\documentclass[fleqn,usenatbib]{mnras}
\usepackage{mathptmx}
\usepackage[T1]{fontenc}
\usepackage{ae,aecompl}
\usepackage{graphicx}
\usepackage{eurosym}
\usepackage{amssymb}

\title[HiPERCAM]{HiPERCAM: a quintuple-beam, high-speed optical imager
  on the 10.4-m Gran Telescopio Canarias}

\author[V. S. Dhillon et al.]{V. S. Dhillon,$^{1,2}$\thanks{E-mail:
    vik.dhillon@sheffield.ac.uk (VSD)} N. Bezawada$^3$, M. Black$^4$,
  S. D. Dixon$^1$, T. Gamble$^1$, X. Gao$^4$,  
\newauthor D. M. Henry$^4$, P. Kerry$^1$, S. P. Littlefair$^1$,
D. W. Lunney$^4$, T. R. Marsh$^5$, 
\newauthor C. Miller$^4$, S. G. Parsons$^1$, R. P. Ashley$^5$, 
E. Breedt$^6$, A. Brown$^1$, M. J. Dyer$^1$, 
\newauthor M. J. Green$^5$, I. Pelisoli$^5$, D. I. Sahman$^1$, J. Wild$^1$,
D. J. Ives$^3$, L. Mehrgan$^3$,
\newauthor J. Stegmeier$^3$, C. M. Dubbeldam$^7$, T. J. Morris$^7$, 
J. Osborn$^7$, R. W. Wilson$^7$, 
\newauthor J. Casares$^{2,8}$, T.  Mu\~{n}oz-Darias$^{2,8}$, E. Pall\'{e}$^{2,8}$, 
P. Rodr\'{i}guez-Gil$^{2,8}$, 
\newauthor T. Shahbaz$^{2,8}$, M. A. P. Torres$^{2,8}$, A. de Ugarte Postigo$^9$,
A. Cabrera-Lavers$^{10,2}$,
\newauthor R. L. M. Corradi$^{10,2}$, R. D. Dom\'{i}nguez$^{10,2}$,
D. Garc\'{i}a-Alvarez$^{10,2}$
\\
$^{1}$Department of Physics and Astronomy, University of Sheffield,
Sheffield S3 7RH, UK \\
$^{2}$Instituto de Astrof\'{i}sica de Canarias, E-38205 La Laguna,
Tenerife, Spain \\
$^{3}$European Southern Observatory, 85748 Garching bei M\"{u}nchen,
Germany \\
$^{4}$UK Astronomy Technology Centre, Royal Observatory Edinburgh,
Edinburgh EH9 3HJ, UK \\
$^{5}$Department of Physics, University of Warwick, Coventry CV4 7AL,
UK \\
$^{6}$Institute of Astronomy, University of Cambridge, Madingley Road,
Cambridge CB3 0HA, UK \\
$^{7}$Department of Physics, University of Durham, Durham DH1 3LE, UK
\\
$^{8}$Departamento de Astrof\'{i}sica, Universidad de La Laguna
s/n, E-38206 La Laguna, Tenerife, Spain \\
$^{9}$Instituto de Astrof\'{i}sica de Andaluc\'{i}a (IAA-CSIC),
Glorieta de la Astronom\'{i}a s/n, E-18008 Granada, Spain \\
$^{10}$GRANTECAN, Cuesta de San Jos\'{e} s/n, E-38712 Bre\~{n}a Baja,
La Palma, Spain \\
}
\date{Accepted 2021 July 21.}
\pubyear{2020}

\begin{document}
\label{firstpage}
\pagerange{\pageref{firstpage}--\pageref{lastpage}}
\maketitle

\begin{abstract}
HiPERCAM is a portable, quintuple-beam optical imager that saw first
light on the 10.4-m Gran Telescopio Canarias (GTC) in 2018. The
instrument uses re-imaging optics and 4 dichroic beamsplitters to
record $u_s\,g_s\,r_s\,i_s\,z_s$ ($320-1060$\,nm) images
simultaneously on its five CCD cameras, each of 3.1\,arcmin (diagonal)
field of view. The detectors in HiPERCAM are frame-transfer devices
cooled thermo-electrically to 183\,K, thereby allowing both
long-exposure, deep imaging of faint targets, as well as high-speed
(over 1000 windowed frames per second) imaging of rapidly varying
targets. A comparison-star pick-off system in the telescope focal
plane increases the effective field of view to 6.7\,arcmin for
differential photometry.  Combining HiPERCAM with the world's largest
optical telescope enables the detection of astronomical sources to
$g_s \sim 23$ in 1\,s and $g_s \sim 28$ in 1\,h. In this paper we
describe the scientific motivation behind HiPERCAM, present its
design, report on its measured performance, and outline some planned
enhancements.
\end{abstract}

\begin{keywords}
instrumentation: detectors -- instrumentation: photometers --
techniques: photometric.
\end{keywords}

\section{Introduction}


The advent of powerful time-domain survey facilities, such as the
Zwicky Transient Facility (ZTF; \citealt{2019PASP..131g8001G}), the
Vera Rubin Observatory (VRO; \citealt{2009arXiv0912.0201L}) and the
Gravitational-wave Optical Transient Observer (GOTO;
\citealt{2018SPIE10704E..0CD}), will revolutionise our knowledge of
the Universe in the coming decades. Detailed follow-up observations of
the most interesting objects discovered by such surveys will be
essential if we are to understand the astrophysics of the
sources. Although the largest telescopes in the world do provide
instrumentation for such follow-up work, one area is poorly
catered for -- high-speed (seconds to milliseconds) optical cameras.

High time resolution probes allows one to test fundamental
physics by probing the most extreme cosmic environments -- black
holes, neutron stars and white dwarfs. For example, neutron stars and
black holes allow the effects of strong-field general relativity to be
studied, and white dwarfs and neutron stars provide us with the
opportunity to study exotic states of matter predicted by quantum
mechanics (e.g. \citealt{2013Sci...340..448A}). White dwarfs, neutron
stars and black holes are also a fossil record of stellar evolution,
and the evolution of such objects within binaries is responsible for
some of the Universe's most exotic phenomena, such as short gamma-ray
bursts, millisecond pulsar binaries, type Ia supernovae, and possibly
fast radio bursts (FRBs; e.g. \citealt{2020ApJ...895L..30L}).

One way of studying compact objects is through their photometric
variability in multiple colours. The dynamical timescales of white
dwarfs, neutron stars and black holes range from seconds to
milliseconds, and hence the pulsation and rotation of these objects,
and the motion of any material in close proximity to them (e.g. in an
accretion disc), tends to occur on these short timescales. Therefore,  
the variability of compact objects can only be resolved by observing
at high speeds, providing information on their masses, radii,
internal structures and emission mechanisms
(e.g. \citealt{2017MNRAS.470.4473P,2017NatAs...1..859G}).

Observing the Universe on timescales of seconds to milliseconds is
also of benefit when studying less massive compact objects, such as
brown dwarfs, exoplanets, and solar system objects. Although the
eclipses and transits of exoplanets occur on timescales of minutes to
hours, observing them at high time resolution can significantly
improve throughput due to the avoidance of detector readout time, and
enables the detection of Earth-mass planets through small variations
in transit timing. By observing in multiple colours simultaneously,
transit light curves of exoplanets are also sensitive to
wavelength-dependent opacity sources in their atmospheres
(e.g. \citealt{2016MNRAS.463.2922K}). High time-resolution occultation
observations of solar system objects enable their shapes and sizes to
be measured, and allow one to detect atmospheres, satellites and ring
systems at spatial scales (0.0005\,arcsec) only otherwise achievable
from dedicated space missions (see \citealt{2012Natur.491..566O}).

In this paper, we describe a new high-speed camera called
HiPERCAM\footnote{http://www.vikdhillon.staff.shef.ac.uk/hipercam},
for High PERformance CAMera, mounted on the world's largest optical
telescope -- the 10.4-m Gran Telescopio Canarias (GTC) on La Palma.
HiPERCAM has been designed to study compact objects of all classes,
including white dwarfs, neutron stars, black holes, brown dwarfs,
exoplanets and the minor bodies of the solar system. However, HiPERCAM
is much more than just a high-speed camera -- it can equally be used
for deep imaging of extended extragalactic targets simultaneously in
five optical colours, making it an extremely efficient general-purpose
optical imager for the GTC. Brief descriptions of the instrument
during the early design and commissioning phases have been provided by
\cite{2016SPIE.9908E..0YD}, \cite{2018SPIE10702E..0LD} and
\cite{2018SPIE10709E..24B}, but no detailed description of the
final instrument has appeared in the refereed astronomical literature -- a
situation rectified by this paper.

\section{Design}

HiPERCAM was designed to be a significant advance upon its
predecessor,
ULTRACAM\footnote{http://www.vikdhillon.staff.shef.ac.uk/ultracam}
\citep{2007MNRAS.378..825D}. ULTRACAM saw first light in 2002 and has
since been used for nearly 700 nights on the 4.2-m William Herschel
Telescope (WHT) on La Palma, the 8.2-m Very Large Telescope (VLT) at
Paranal, and the 3.5-m New Technology Telescope (NTT) on La Silla,
where it is now permanently mounted. Some of the scientific highlights
of ULTRACAM include the discovery of brown-dwarf mass donors in
cataclysmic variables \citep{2006Sci...314.1578L}, discovery of the
first white-dwarf pulsar \citep{2016Natur.537..374M}, and measurement
of the size and albedo of the dwarf planet Makemake
\citep{2012Natur.491..566O}.

The HiPERCAM project began in 2014 and saw first light four years
later as a visitor instrument on the GTC, on budget (\euro3.5M) and on
time. HiPERCAM's performance far exceeds that of ULTRACAM. HiPERCAM
can image simultaneously in 5 SDSS (Sloan Digital Sky Survey) bands
($ugriz$) rather than the 3 bands of ULTRACAM ($ugr$, $ugi$ or
$ugz$). HiPERCAM can frame at windowed rates of over 1 kHz,
rather than the few hundred Hz of ULTRACAM. HiPERCAM uses detectors
cooled to 183\,K compared to the 233\,K of ULTRACAM, with
deep-depletion, anti-etalon CCDs in the red channels (see
Section~\ref{sec:detectors}), resulting in much lower dark current,
higher quantum efficiency and lower fringing than those in
ULTRACAM. Hence, although designed for high-speed observations,
HiPERCAM is also well-suited to science programs that require deep
(i.e. long exposure), single-shot spectral-energy distributions, such
as light curves of extragalactic transients
(e.g. \citealt{2018NatAs...2..751L}) and stellar population studies of
low surface-brightness galaxies
(e.g. \citealt{2016ApJ...823..123T}). HiPERCAM also has twice the
field of view of ULTRACAM (when mounted on the same telescope) and a
novel comparison-star pick-off system, providing more comparison stars
for differential photometry of bright targets, such as the host stars
of transiting exoplanets. Each of these design improvements are
described in greater detail below.

\subsection{Optics}
\label{sec:optics}

Like ULTRACAM, HiPERCAM was originally designed to be a visitor
instrument, moving between 4--10\,m-class telescopes in both
hemispheres to maximise its scientific potential. Hence, the baseline
optical design for HiPERCAM was optimised to provide good imaging
performance on the WHT, NTT and GTC.

\begin{figure*}
  \includegraphics[width=8.6cm]{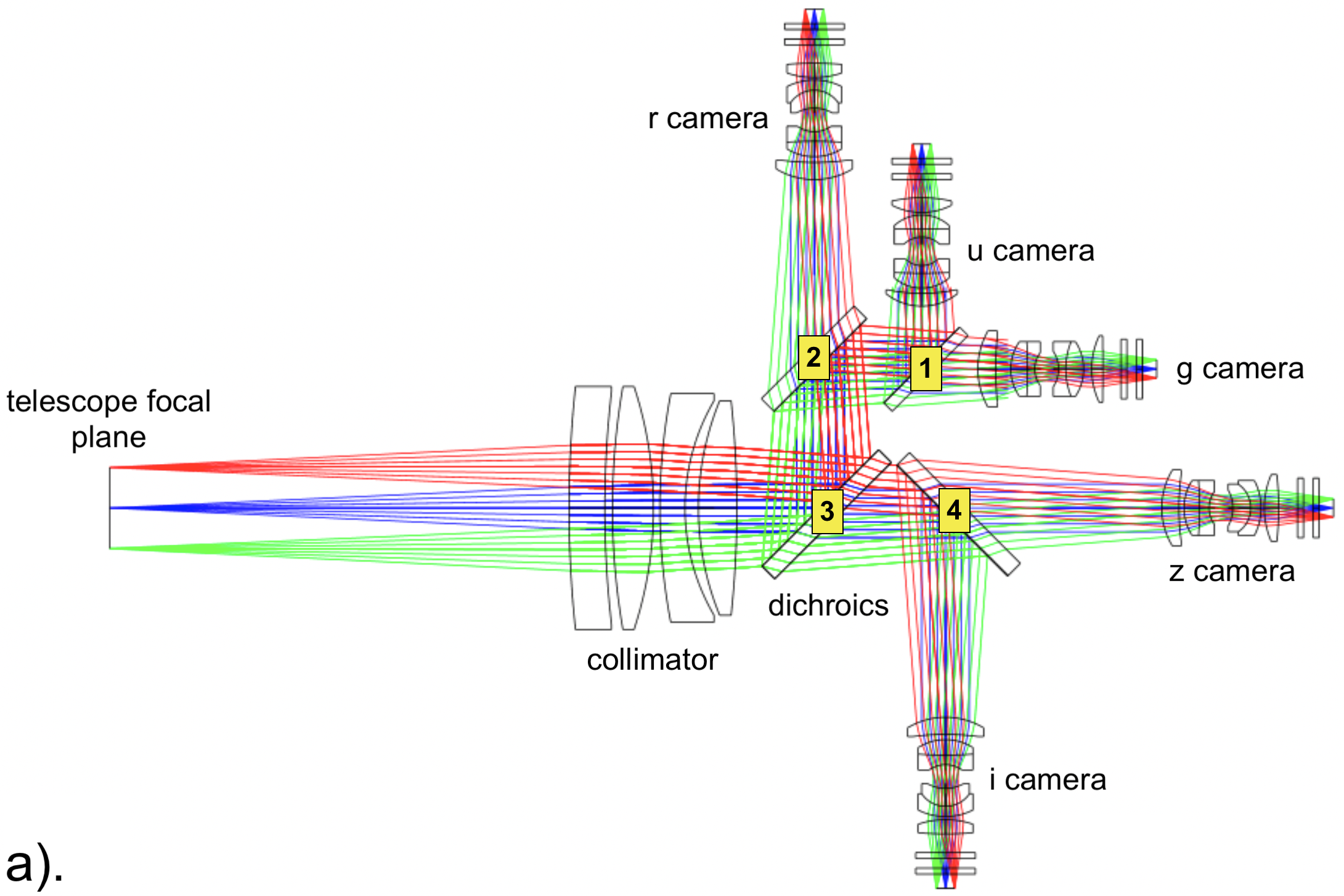}~\includegraphics[width=9.0cm]{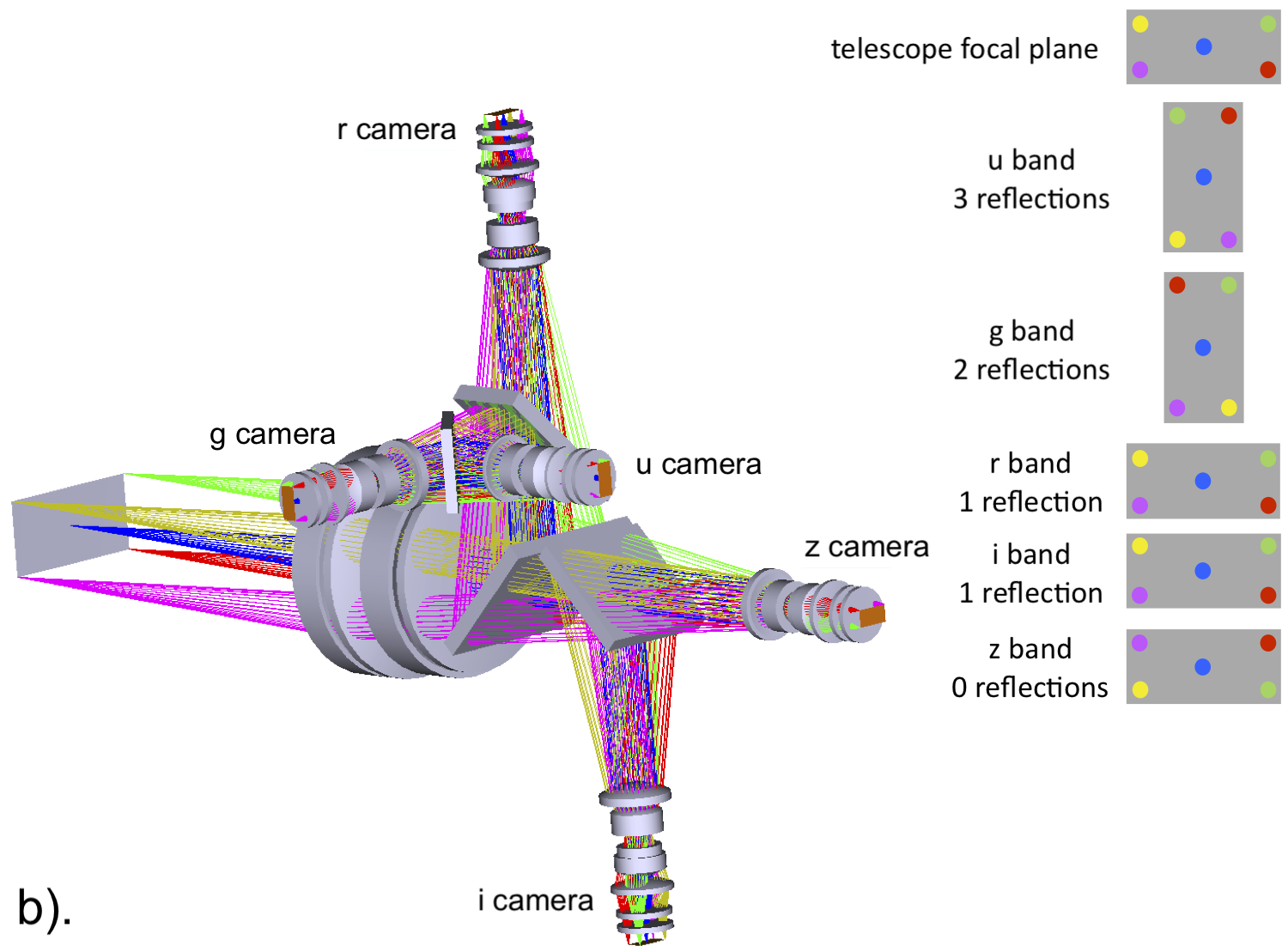}
\caption{a). Ray trace through the HiPERCAM optics. The red, green and
  blue lines represent ray bundles emanating from three
  spatially-separated point sources in the telescope focal plane. The
  diagram is to scale -- for reference, the diameter of the largest
  lens in the collimator is 219\,mm. The yellow boxes show the
  dichroic numbers, which are in ascending order of the cut-point
  wavelength, as shown in
  \protect{Figure~\ref{fig:filters}}. b). Three-dimensional view of
  the HiPERCAM optics, showing the dichroic rotations used to package
  the instrument more efficiently. The orientation of each detector
  with respect to the telescope focal plane is shown on the far right,
  with the coloured spots corresponding to the bundles of rays of the
  same colour shown in the three-dimensional view.}
\label{fig:raytrace}
\end{figure*}

\subsubsection{Requirements}
\label{sec:reqts}

The main requirements of the HiPERCAM optics were as follows:

\begin{enumerate}  
\item To provide simultaneous imaging in 5 optical bands covering the
  SDSS $ugriz$ filter bandpasses from 320--1000\,nm
  \citep{1996AJ....111.1748F}.
\item To give a plate scale of 0.3\,arcsec/pixel on the WHT. Using
  an e2v 231-42 CCD with 2048$\times$1024 imaging pixels, each of
  15\,$\mu$m in size (see Section~\ref{sec:detectors}), this plate
  scale would provide a field of view of 10.24$\times$5.12\,arcmin on
  the WHT.
\item The optics should not degrade the point-spread-function (PSF) by
  more than 10 per cent, measured over 80 per cent (radius) of the
  field of view in median WHT seeing conditions (see
  \citealt{1999MNRAS.309..379W}) of 0.68/0.64/0.61/0.58/0.56\,arcsec
  in $ugriz$. Hence in seeing of 0.68\,arcsec in the $u$
  band, for example, the stellar PSFs should have FWHM (full width at
  half maximum) of $<0.75$\,arcsec out to a field
  radius of 4.5\,arcmin.
\item The plate scale should be constant across this wavelength range
  to within 0.0005\,arcsec/pixel, thereby ensuring that stars have the
  same CCD pixel positions in each band, to within $\pm$\,1 pixel
  (assuming perfect alignment of the CCDs relative to each other).
\item The optical design should be optimised so that the image quality
  of HiPERCAM when used on the NTT and GTC should be equivalent to
  that on the WHT. Using the CCD specified in (ii), the optics
  would provide a plate scale of 0.357 and 0.081\,arcsec/pixel, and a
  field of view of 12.18$\times$6.09\,arcmin and
  2.76$\times$1.38\,arcmin, on the NTT and GTC, respectively.
\end{enumerate}

\subsubsection{Layout}

A ray trace through the HiPERCAM optics is shown in
Figure~\ref{fig:raytrace}a. Light from the telescope focal plane is
first collimated by a four-element collimator. It then passes through
four dichroic beamsplitters that split the light into five
wavebands. Each of the five collimated beams is then focused by a
re-imaging camera onto a detector. The re-imaging cameras are of a
double-Gauss type design with two singlet lenses and two cemented
doublet lenses arranged in a roughly symmetrical layout around the
re-imaged pupil. The light then travels through a bandpass filter and
a cryostat window before striking the detector.

The layout in Figure~\ref{fig:raytrace}a shows all five cameras in the
same plane; this is for clarity only, and in reality a more compact
packaging of the dichroics and associated re-imaging cameras has been
achieved by rotating them around the optical axis of the system, as
shown in Figure~\ref{fig:raytrace}b. As a result of the differing
number of dichroic reflections experienced by the beams, the images
falling on the detectors are flipped with respect to each other (see
the coloured spots at the far right in
Figure~\ref{fig:raytrace}b). This is corrected in the data acquisition
software (see Section~\ref{sec:das}) to ensure that the output images
have the same orientation and left-right/top-bottom flip.

\begin{table*}
\centering
\caption{Summary of the main optical parameters of HiPERCAM on the
  three telescopes for which the optical design was optimised.}
\begin{tabular}{lccc}
\hline\noalign{\smallskip}
 & WHT & NTT & GTC \\
\hline\noalign{\smallskip}
Telescope design & Cassegrain & Ritchey-Chr\'{e}tien & Ritchey-Chr\'{e}tien \\
Entrance pupil diameter (mm) & 4179.0 & 3500.0 & 11053.4$^*$ \\
Effective focal length (mm) & 45737.5 & 38501.7 & 169897.7 \\
Telescope f-ratio & 10.95 & 11.00 & 15.415 \\
Telescope focal-plane scale ($^{\prime\prime}$/mm) & 4.510 & 5.357 & 1.214 \\
Detector plate scale ($^{\prime\prime}$/mm) & 20.0 & 23.759 & 5.382 \\
Detector pixel scale ($^{\prime\prime}$/pixel) & 0.300 & 0.356 & 0.081 \\
Detector field of view ($^{\prime}$) & 10.24$\times$5.12 & 12.16$\times$6.08 & 2.76$\times$1.38 \\
Internal pupil diameter$^{\dagger}$ (mm) & 21.7 & 21.4 & 15.5 \\
f-ratio at detector & 2.468 & 2.480 & 3.465 \\
\hline\noalign{\smallskip}
\multicolumn{4}{l}{
\begin{minipage}[t]{1.4\columnwidth}
$^*$This is the distance across the segmented, hexagonal primary
  mirror. The diameter of a circular mirror with the same collecting
  area as the GTC primary would be 10.4\,m.
\end{minipage}
} \\
\noalign{\smallskip}
\multicolumn{4}{l}{
\begin{minipage}[t]{1.4\columnwidth}
$^{\dagger}$This is the diameter of the intermediate pupil lying
  within the re-imaging cameras that is conjugate with the entrance
  aperture of the telescope. The entrance aperture of the GTC lies at
  the secondary mirror, whereas on the WHT and NTT it is the primary
  mirror.
\end{minipage}
} \\
\end{tabular}
\label{tab:optics}
\end{table*}

The HiPERCAM collimator and re-imaging cameras together operate as a
focal reducer, demagnifying the image in the telescope focal plane by
a factor of 0.225, given by the ratio of the re-imaging camera focal
length (98.6\,mm) to the collimator focal length (437.3\,mm). A
summary of the main optical parameters of HiPERCAM on the three
telescopes for which the optical design was optimised is given in
Table~\ref{tab:optics}. Note that an optical design does exist for a
separate GTC collimator which, with no change to any of the other
HiPERCAM optics, would increase the detector pixel scale and field of
view to 0.113\,arcsec/pixel and 3.84$\times$1.68\,arcmin,
respectively. However, given its high cost, the excellent image
quality obtained on the GTC with the existing collimator, and the
effective increase in the field of view of HiPERCAM on the GTC
afforded by COMPO (see Section~\ref{sec:compo}), this second
collimator has not been built.

The 4-lens collimator is the first optical component and hence must have
high transmission across the required 320--1000\,nm wavelength range.
The glasses chosen were therefore N-BAK2, CaF$_2$ and LLF1, with the largest
lens of diameter 219\,mm. The 6-element re-imaging cameras for the
three longer wavebands ($riz$) share a common optical design, but the
$ug$ camera designs are unique in order to maximise throughput and
image quality. The first element in each re-imaging camera was
manufactured last, to re-optimised radii of curvature and thicknesses
based on the as-built properties of the other five lenses.  This
compensated for differences in glass dispersion and tolerance
build-up, thereby ensuring that all bands have the same effective
pixel scale and optimum image quality. The lens-lens axial spacings
were also re-optimised during this process. All lenses were
anti-reflection coated, with the collimator lenses receiving a
broadband coating with average reflectivity of $<$\,2 per cent, and
the re-imaging lenses receiving a narrow-band coating with average
reflectivity $<$\,0.5 per cent. The HiPERCAM lenses were manufactured
by the Rocky Mountain Instrument Company, Colorado, who also performed
the anti-reflection coating and mounted the lenses in aluminium
barrels (see Section~\ref{sec:mech}).

\begin{figure}
\includegraphics[width=8.6cm]{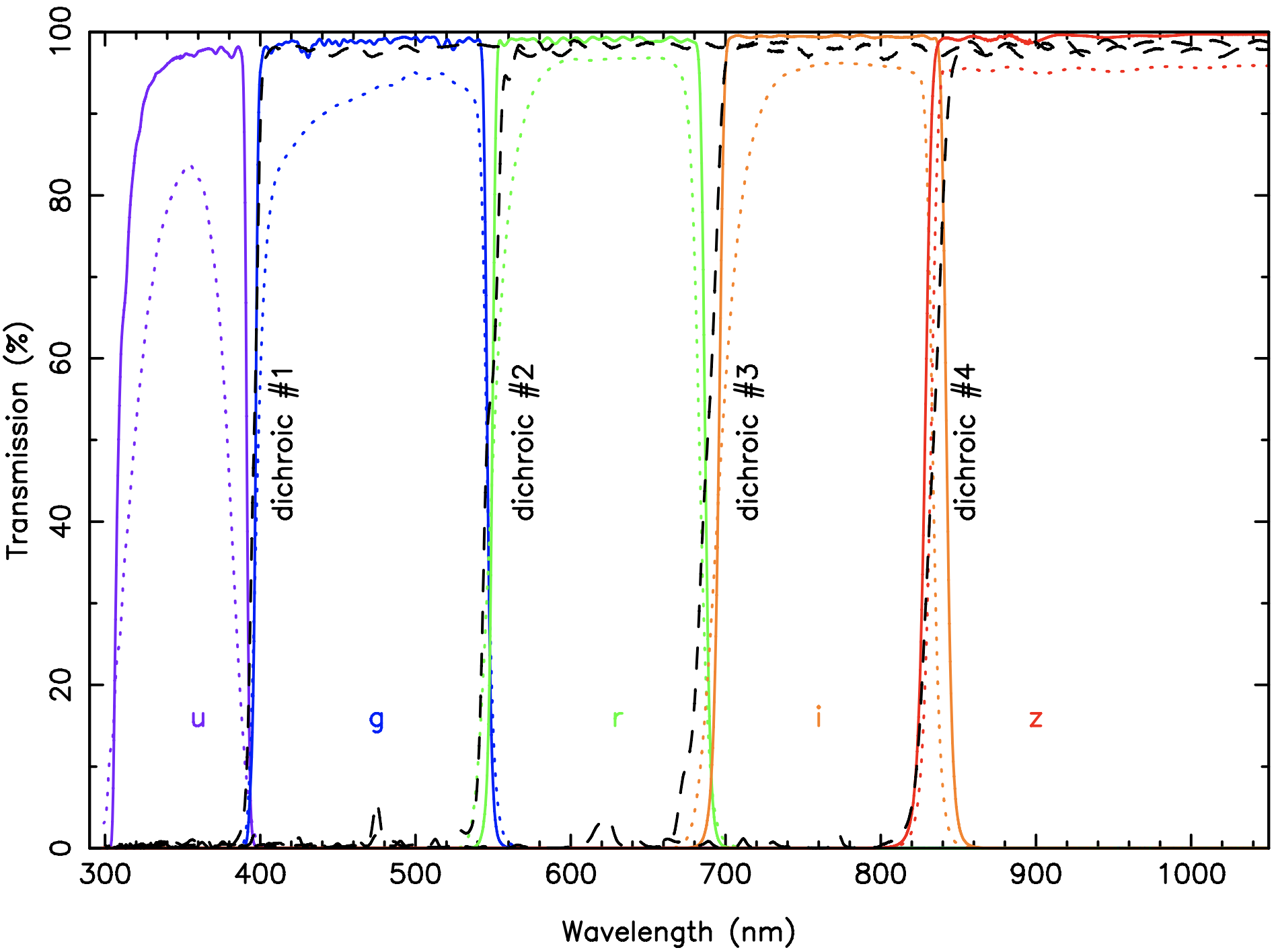}
\caption{Transmission profiles of the as-built HiPERCAM dichroic
  beamsplitters (dashed lines), the HiPERCAM ``original'' SDSS filters
  (dotted lines), and the HiPERCAM ``Super'' SDSS filters (solid
  lines). One of the main advantages of HiPERCAM over its predecessor,
  ULTRACAM, is that one no longer has to choose which of $riz$ to
  select for the red arm filter, as all three are simultaneously
  available.}
\label{fig:filters}
\end{figure}

The four dichroic beamsplitters are made of fused silica, with the
largest of dimension 140$\times$150\,mm. The front faces of the
dichroics are coated with long-wave pass coatings that reflect
incident light with wavelengths shorter than the cut-points and
transmit longer wavelengths. The dichroic cut points are shown in
Figure~\ref{fig:filters} and were selected to maximise the throughput
in the two adjacent filter bandpasses. This calculation was performed
after the filters had been manufactured and hence their as-built
bandpasses were known. The difference between the wavelengths at which
90 per cent and 10 per cent transmission occurs is $<$\,15.5\,nm, and
the reflectance/transmission of wavelengths shorter/longer than the
cut points is $>$\,99.5 per cent and $>$\,98 per cent,
respectively. To maximise throughput and minimise ghosting, the rear
of each dichroic is coated with a narrow-band anti-reflection coating
of average reflectivity $<$\,0.5 per cent. Detailed modelling of the
ghosting in the HiPERCAM optics showed that the brightest ghosts are
$\sim 10^7$ times fainter than the primary image, which is
insignificant given that the dynamic range of the detector is of order
10$^4$.

The bandpasses of HiPERCAM's five arms are defined by a set of
so-called ``Super'' SDSS filters (Figure~\ref{fig:filters}). These
filters, which we refer to as $u_sg_sr_si_sz_s$, were specifically
designed for HiPERCAM, with cut-on/off wavelengths that match the
original SDSS $ugriz$ filters \citep{1996AJ....111.1748F} but which
use multi-layer coatings rather than coloured glasses to define the
bandpasses and increase throughput. The $u_sg_sr_si_sz_s$ filters
provide a throughput improvement of 41/9/6/9/5 per cent compared to
$ugriz$ filters, respectively. Since the whole optical spectrum is
being covered in one shot by HiPERCAM, we decided not to install
filter wheels in front of each CCD. Instead, each filter is mounted in
a cartridge which can be easily changed by hand, if required. The
HiPERCAM filters are of identical dimensions (50$\times$50\,mm) and
optical thicknesses to the ULTRACAM filters, which means that the
extensive set of ULTRACAM narrow-band absorption-line, emission-line
and continuum
filters\footnote{http://www.vikdhillon.staff.shef.ac.uk/ultracam/filters/filters.html}
can be used in HiPERCAM.

The final element in the optical path of each HiPERCAM arm is a
fused-silica window, which allows light onto the CCD whilst forming a
vacuum seal with the detector head (see
Section~\ref{sec:detectors}). The windows have broadband
anti-reflection coatings with average reflectivity of $<$\,1 per
cent. The HiPERCAM dichroics, filters and windows were all
manufactured by Asahi Spectra Company, Tokyo. The HiPERCAM optics are
far superior in terms of throughput and image quality compared to
ULTRACAM, having benefitted from a ten-fold increase in the optics
budget and nearly two decades of improvements in optical manufacturing
and coating techniques.

\section{Mechanical design}
\label{sec:mech}

\begin{figure*}
\centerline{\includegraphics[width=15.0cm]{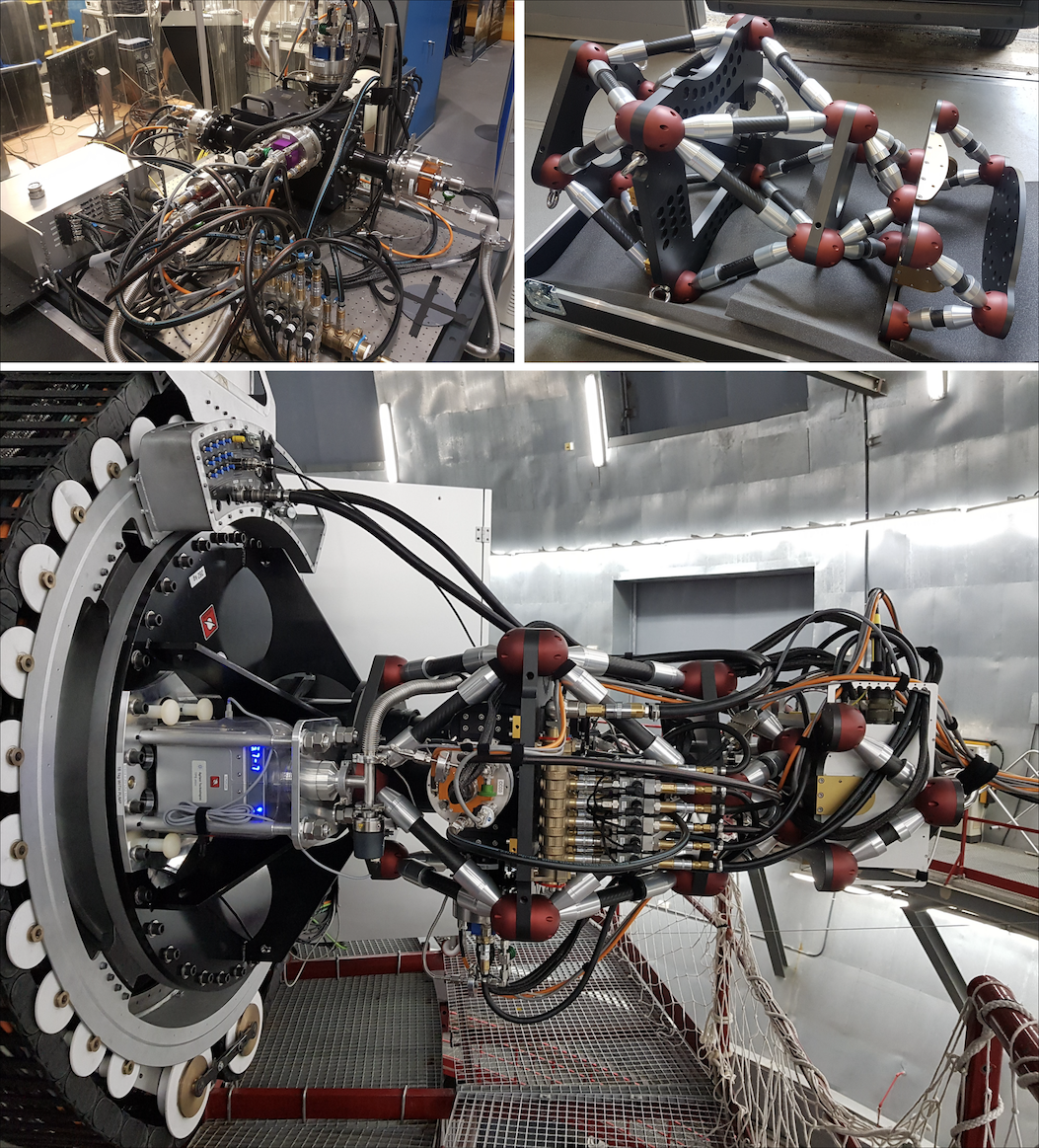}}
\caption{{\em Top left:} The HiPERCAM hull during alignment at the UK
  Astronomy Technology Centre (UKATC).  The hull is the black box at
  the centre of the image. The five re-imaging cameras and CCD heads
  can be seen attached to the hull. The rears of the CCD heads are
  anodised according to the filter colour ($u_s$ = violet, $g_s$ =
  blue, $r_s$ = orange, $i_s$ = red, $z_s$ = dark red) for ease of
  identification. The aluminium box at the lower left is the CCD
  controller. {\em Top right:} The HiPERCAM opto-mechanical chassis
  during integration at the UKATC. From left to right, the first three
  vertical black plates are the top plate (which attaches to the
  telescope), the middle plate (to which the hull is attached) and the
  bottom plate (to which the cradle holding the CCD controller is
  attached). For scale, the total length of the opto-mechanical
  chassis is 1.25\,m. {\em Bottom:} HiPERCAM on the Folded Cassegrain
  focus of the GTC. From left to right can be seen the rotator
  (surrounded by its cable wrap), the black HiPERCAM interface collar
  (on which is mounted the vacuum pump with blue LEDs in the image),
  and HiPERCAM.}
\label{fig:mech}
\end{figure*}

The mechanical structure of HiPERCAM was designed to meet the
following requirements:

\begin{enumerate}
\item Provide a rigid platform on which to mount the optics and CCD
  heads, with relative flexure between the CCD heads of less than
  $\sim$1\,pixel (15\,$\mu$m) at any instrument orientation, so that
  stars do not drift out of small windows defined on the five CCDs.
\item Provide a mounting for the CCD controller, which must lie within
  a cable length of $\sim$1.5\,m of the CCD heads to
  minimise pickup noise and clock-signal degradation.
\item Allow access to the CCD heads for maintenance, alignment and
  filter changes.
\item Minimise thermal expansion for focus stability.
\item Provide electrical and thermal isolation from the telescope to
  reduce pickup noise via ground loops and minimise the load on the
  water cooling system.
\item Provide a light-tight and dust-proof environment for the optics.
\item Have a total weight of $<$\,1000\,kg, set by the mass limit of the
  GTC rotator, and size $<$\,$1.3\times1.0\times1.0$\,m, set by the
  intersection of the GTC instrument space envelope at the Folded
  Cassegrain focus and the maximum dimensions of a single item that can be
  air freighted to La Palma.
\end{enumerate}

To meet the above requirements, the HiPERCAM opto-mechanical chassis
is composed of 3 aluminium plates connected by carbon fibre
struts. This triple-octopod design is shown in Figure~\ref{fig:mech}
and provides an open, stiff, compact (1.25\,m long) and light-weight
(288\,kg) structure that is relatively insensitive to temperature
fluctuations. These characteristics also make it straight-forward to
transport, maintain and mount/dismount HiPERCAM at the telescope. The
collimator, dichroics, re-imaging lenses, filters and CCD heads are
all housed in/on an aluminium hull that forms a sealed system to dust
and light. The hull is attached to the central aluminium plate, the
CCD controller is mounted in a cradle hanging off the bottom plate,
and the top plate connects the instrument to the telescope, as shown
in Figure~\ref{fig:mech}. A steel interface collar attaches HiPERCAM
to the rotator and places the instrument at the correct back-focal
distance. A layer of G10/40 epoxy glass laminate is located between
the top plate of HiPERCAM and the collar to provide electrical and
thermal isolation from the telescope. The mounting collar houses a
motorised focal-plane mask. This is an aluminium blade that can be
inserted in the telescope focal plane to block light from bright stars
falling on the active area of the sensor above the CCD windows, which
would otherwise cause vertical streaks in the images. This mask also
prevents photons from stars and the sky from contaminating the windows
in drift-mode (see Section~\ref{sec:modes}).

\subsection{Detectors}
\label{sec:detectors}

\begin{figure*}
\centerline{\includegraphics[width=15.0cm]{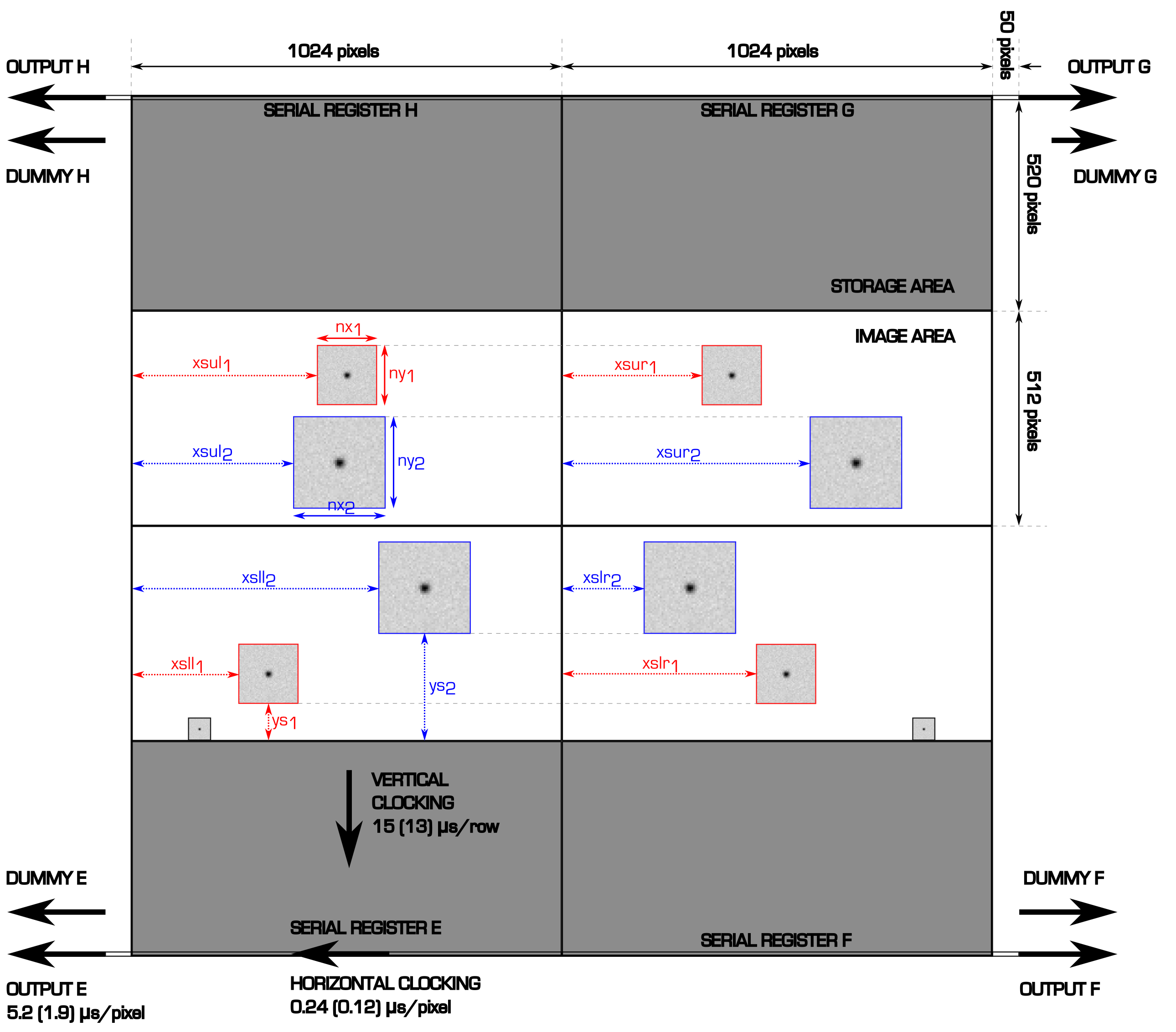}}
\caption{A schematic of the CCD231-42 detector used in HiPERCAM. The
  CCD has a split frame-transfer architecture with four outputs,
  labelled E, F, G and H by e2v, and four dummy outputs for
  common-mode signal rejection. The image area is shown in white and
  the storage area in grey. The lower-left quadrant is read by output
  E, the lower right by F, the upper right by G and the upper left by
  H. There are four 1024-pixel serial registers, two at the top and
  two at the bottom of the detector, which can be clocked
  independently and which have an additional 50 pre-scan pixels for
  bias-level determination. The storage area is 8 pixels larger in the
  vertical direction than the image area, and these over-scan pixels
  can also be used to determine the bias level. The pixel and clocking rates
  indicated in the diagram are for the slow settings -- values for the
  fast settings are given in brackets. The detector can be read out in
  three different modes: 1. Full-frame mode, where the entire white
  region is read out; 2. Windowed mode, where either the four red
  windows (one ``quad'') are read out, or the four red {\em and} the
  four blue windows are read out (two quads); 3. Drift mode, where the
  two small black windows on the border between the lower image and
  storage areas are read out.}
\label{fig:ccd}
\end{figure*}

%

HiPERCAM employs 5 custom-designed CCD231-42 detectors from Teledyne
e2v. The CCDs are split frame-transfer devices with 15\,$\mu$m pixels
and 4 outputs, with one output located at each corner. The devices
have a format of 2048$\times$2080 pixels, where the upper
2048$\times$520 and lower 2048$\times$520 pixels are coated with
reflective aluminium masks and used as frame-transfer storage areas,
providing a central image area of 2048$\times$1024 pixels. Each CCD
output therefore processes a quadrant of 1024$\times$512 pixels, as
shown in Figure~\ref{fig:ccd}.

\begin{figure}
\centerline{\includegraphics[width=8.6cm]{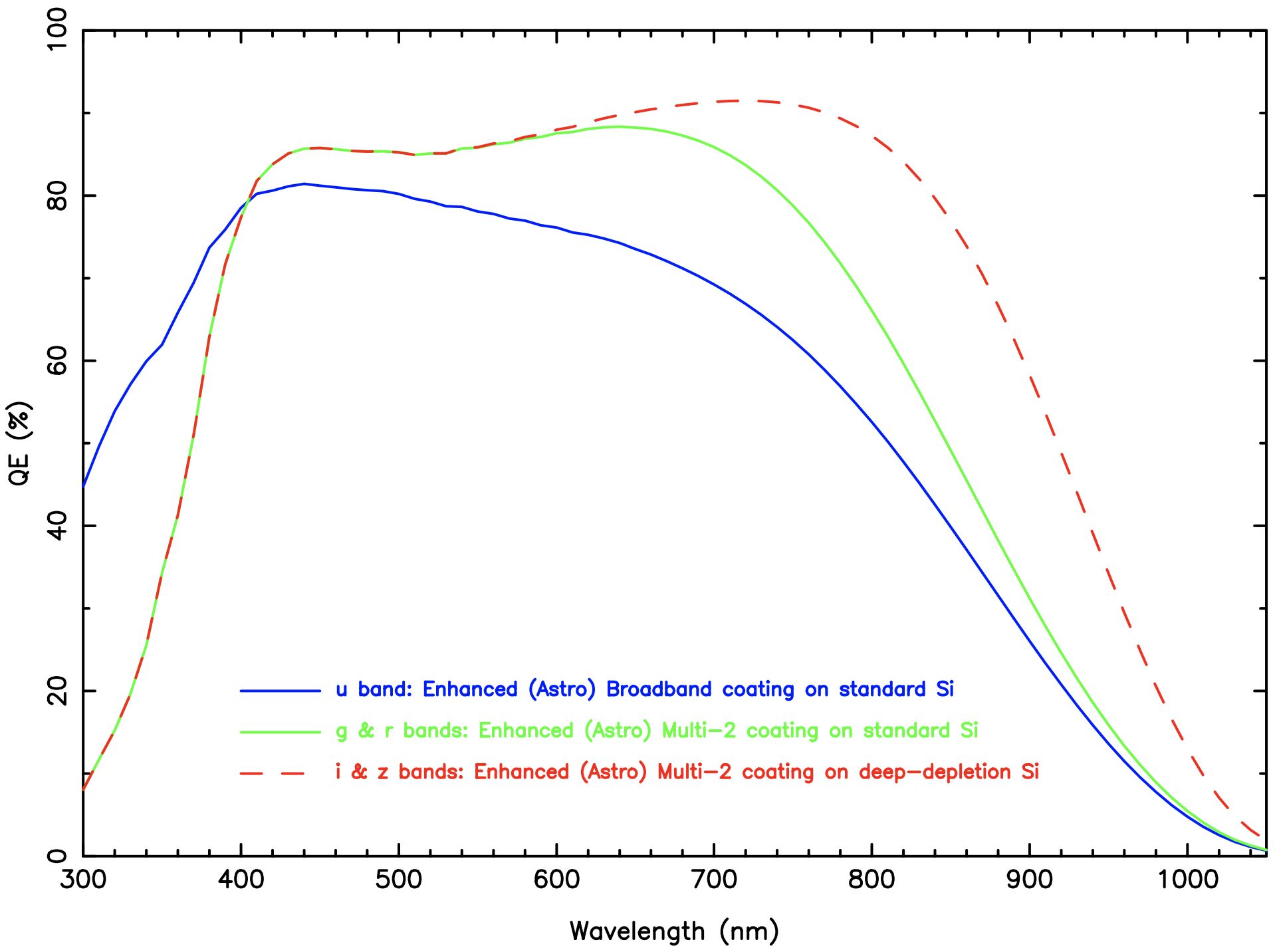}}
\caption{Quantum efficiency curves of the HiPERCAM CCDs at 173\,K.}
\label{fig:qe}
\end{figure}

The CCDs used in HiPERCAM are back-illuminated and thinned to maximise
quantum efficiency (QE) -- the QE curves are shown in
Figure~\ref{fig:qe}. All of the CCDs are anti-reflection (AR) coated
-- the $u_s$-band CCD with the Enhanced (or ``Astro'') Broadband
coating, and the $g_sr_si_sz_s$ CCDs with the Astro Multi-2
coating. The $u_sg_sr_s$ CCDs are manufactured from standard silicon
and, to maximise red QE, the $i_sz_s$ CCDs are manufactured from
deep-depletion silicon. The $i_sz_s$ CCDs have also undergone e2v's
fringe suppression (anti-etaloning) process, where irregularities in
the surface of the CCD are introduced to break the interference
condition. This reduces the $i_s$-band fringing to essentially zero
and the $z_s$-band fringing to approximately the same level as the
$\sim1$ per cent flat-field noise (see \citealt{tulloch95}). The
HiPERCAM CCDs are of the highest cosmetic quality available (grade 1)
and have a full-well capacity of $\sim$115\,000\,e$^-$. The CCDs
are operated with a system gain of 1.2\,e$^-$/ADU and 16-bit
analogue-to-digital converters (ADCs) in the CCD controller (see
Section~\ref{sec:das}), thereby adequately sampling the read noise to
minimise quantisation noise, and ensuring a reasonable match between
digital saturation and device saturation.

\begin{figure*}
\centerline{\includegraphics[width=7cm]{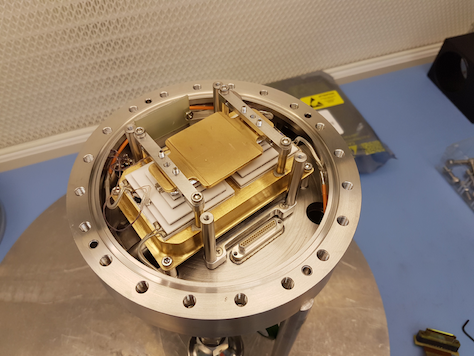}~\includegraphics[width=7cm]{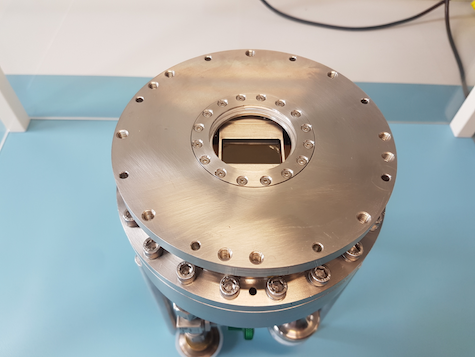}~\includegraphics[height=5.25cm]{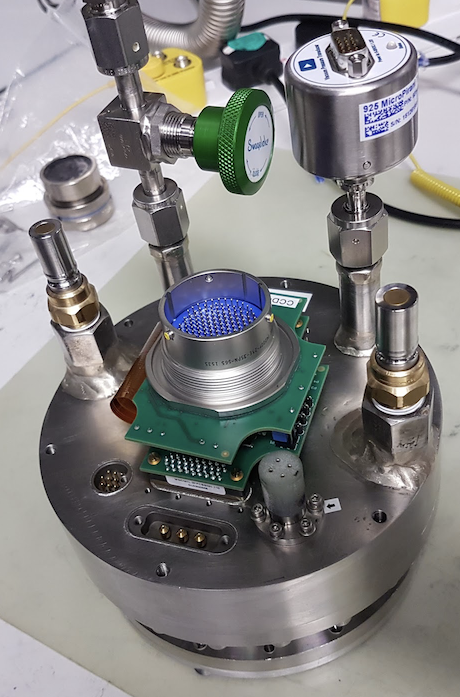}}
\caption{{\em Left:} A view of the interior of one of the HiPERCAM CCD
  heads. The gold-plated cold plate on which the CCD is mounted sits
  on top of two, white 5-stage TECs. The two TECs sit side-by-side on
  a gold-plated heatsink through which the coolant flows. {\em
    Centre:} Front view of the head, showing the CCD through the
  window. The weight of the head is approximately 7\,kg and its
  diameter is 160\,mm. {\em Right:} Rear view of the head, showing the
  green pre-amplifier board with blue 128-pin connector, the green
  vacuum valve, the vacuum gauge, the quick-release connectors for the
  coolant inlet/outlet, and the connector sockets for the temperature
  sensors, TEC power and getter. In this photograph, the
  colour-anodised aluminium box that provides
  electromagnetic-interference shielding of the pre-amplifier board
  (as shown in Figure~\ref{fig:mech}) has been removed.}
\label{fig:head}
\end{figure*}

To minimise read noise and maximise readout speed, the CCDs used in
HiPERCAM are equipped with: low noise amplifiers of 3.2\,e$^-$ rms at
200\,kHz pixel rates (as measured by e2v; see Section~\ref{sec:rno}
for readout-noise measurements at the telescope); dummy outputs to
eliminate pickup noise; fast serial (horizontal) and vertical
(parallel) clocking -- see Figure~\ref{fig:ccd} for rates -- whilst
retaining CTE (charge-transfer efficiency) of $>$\,99.9995 per cent;
independent clocking of the serial register in each quadrant to
provide efficient windowing modes (see Section~\ref{sec:modes});
two-phase image and storage clocks to minimise the frame-transfer
time.

HiPERCAM uses non inverted-mode (NIMO) instead of advanced
inverted-mode (AIMO) CCDs. There are four reasons for this. First, it
is possible to clock NIMO devices more quickly. Second, NIMO devices
have greater well depths. Third, although both NIMO and AIMO CCDs can
have the same dark current specifications at their optimum operating
temperatures, our experience with ULTRASPEC (NIMO,
\citealt{2014MNRAS.444.4009D}) and ULTRACAM (AIMO,
\citealt{2007MNRAS.378..825D}) is that the dark current in NIMO
devices is evenly distributed across the CCD whereas the dark current
in AIMO CCDs is in the form of hot pixels which do not subtract well
using dark frames, making exposures $\gtrsim 30$\,s
undesirable. Fourth, we chose to use NIMO devices in HiPERCAM because
the red CCDs are made of deep-depletion silicon to maximise QE and
this is not compatible with inverted-mode operation.

One consequence of selecting NIMO devices for HiPERCAM is that the
CCDs require cooling to below 187\,K to reduce the dark current to
less than 360\,e$^-$/pixel/h, corresponding to 10 per cent of the
faintest sky level recorded by HiPERCAM (given by $u_s$-band
observations in dark time on the GTC). Therefore, cooling to below
187\,K ensures that dark current is always a negligible noise source
in HiPERCAM. We considered a number of cooling options to meet this
temperature requirement. Liquid nitrogen was rejected as being
impractical -- five liquid-nitrogen cryostats would make HiPERCAM
heavy, large, and time-consuming to fill each night, and continuous
flow or automatic filling systems are not viable given that HiPERCAM
was designed as a visitor instrument with no requirement for dedicated
infrastructure at the telescope. We also rejected closed-cycle
Joule-Thomson coolers, such as the CryoTiger, as it would be difficult
to pass 10 stainless-steel braided gas lines through the cable wrap
and accommodate the 5 compressors at the telescope. Stirling coolers
were given serious consideration, but we were concerned about the
impact of their vibrations on the image quality at the
telescope. Although the vibrations can be reduced, e.g. through the
use of complex, bulky anti-vibration mounts
\citep{2013arXiv1311.0685R}, even with such precautions in place it
would have been difficult to persuade the potential host telescopes to
accept HiPERCAM as a visitor instrument due to the residual
vibrations. Finally, after extensive prototyping and testing to verify
that they could achieve the required CCD temperature, we selected
thermo-electric (Peltier) coolers (TECs), as they are the cheapest,
simplest, lightest and most compact of all of the cooling options.

Our cooling solution, implemented by Universal Cryogenics, Tucson,
uses two side-by-side Marlow NL5010 five-stage TECs, as shown in
Figure~\ref{fig:head}. The detector heads are manufactured from
stainless steel and use all-metal seals rather than rubber o-rings in
order to minimise vacuum leaks. Wherever possible, we avoided the use
of materials that could outgas inside the detectors heads. So, for
example: the pre-amplifier boards were mounted outside the heads (see
\citealt{2018SPIE10709E..24B} for details); corrugated indium foil was
used for the thermal connections between the cold plate, TECs and
heatsink; we installed a non-evaporable porous getter in each head
that acts as an internal vacuum pump and can be periodically
re-activated by heating to $500^{\circ}$C using an external power
supply. Outgassing was further minimised by cleaning the components
ultrasonically prior to assembly, and baking the assembled head whilst
vacuum pumping. Even with these precautions, the vacuum hold time of
the HiPERCAM CCD heads at pressures below $\sim 10^{-3}$\,mbar is only
of order weeks, due primarily to the lack of a sufficiently cold,
large-area interior surface to give effective cryopumping, and
residual outgassing in the small interior volume ($\sim0.5$\,litre) of
the heads. The low-volume heads do, however, have the advantage of
requiring only a few minutes of pumping to bring them back down to
their operating pressure using a 5-way vacuum manifold circuit
permanently installed on HiPERCAM.

A copper heatsink connected to the 280\,K water-glycol cooling circuit
at the GTC is used to extract the heat generated by the TECs in each
CCD head. The heatsinks in the 5 CCD heads are connected in parallel
using two 6-way manifolds (with the sixth channel for cooling the CCD
controller), thereby ensuring that cooling fluid of the same
temperature enters each head. Each of the 6 cooling channels is
equipped with an optical flow sensor made by Titan Enterprises, all of
which are connected to a single Honeywell Minitrend GR Data Recorder
mounted in the electronics cabinet. The data recorder provides a
display of the flow rate in each CCD head and, to protect the CCDs
from overheating, it also has relays to turn off power to the TEC
power supplies if the flow rate in any head drops below a user-defined
limit. The TEC power supplies (made by Meerstetter, model LTR-1200)
have a high-temperature automatic cut-off facility that provides an
emergency backup to the flow sensors: if the temperature of the
heatsink in a CCD head rises above a user-defined value due to a
coolant failure, the power to the TEC is turned off. The TEC power
supplies are able to maintain the HiPERCAM CCD temperatures at their
183\,K set points to within 0.1$^{\circ}$C. At this temperature, the
dark current is only $\sim 20$\,e$^-$/pixel/h.

To prevent condensation on the CCD windows in high humidity
conditions, HiPERCAM employs a 5-way manifold that enables dry, clean
air from the telescope supply to be blown across the outer faces of
the windows at approximately 1\,litre/min. Each CCD head also contains
an internal LED that can be turned on and off for a user-specified
duration to provide a convenient and controllable light source for
testing the detectors.

\subsection{Data acquisition system}
\label{sec:das}

\begin{figure*}
\centerline{\includegraphics[width=15.0cm]{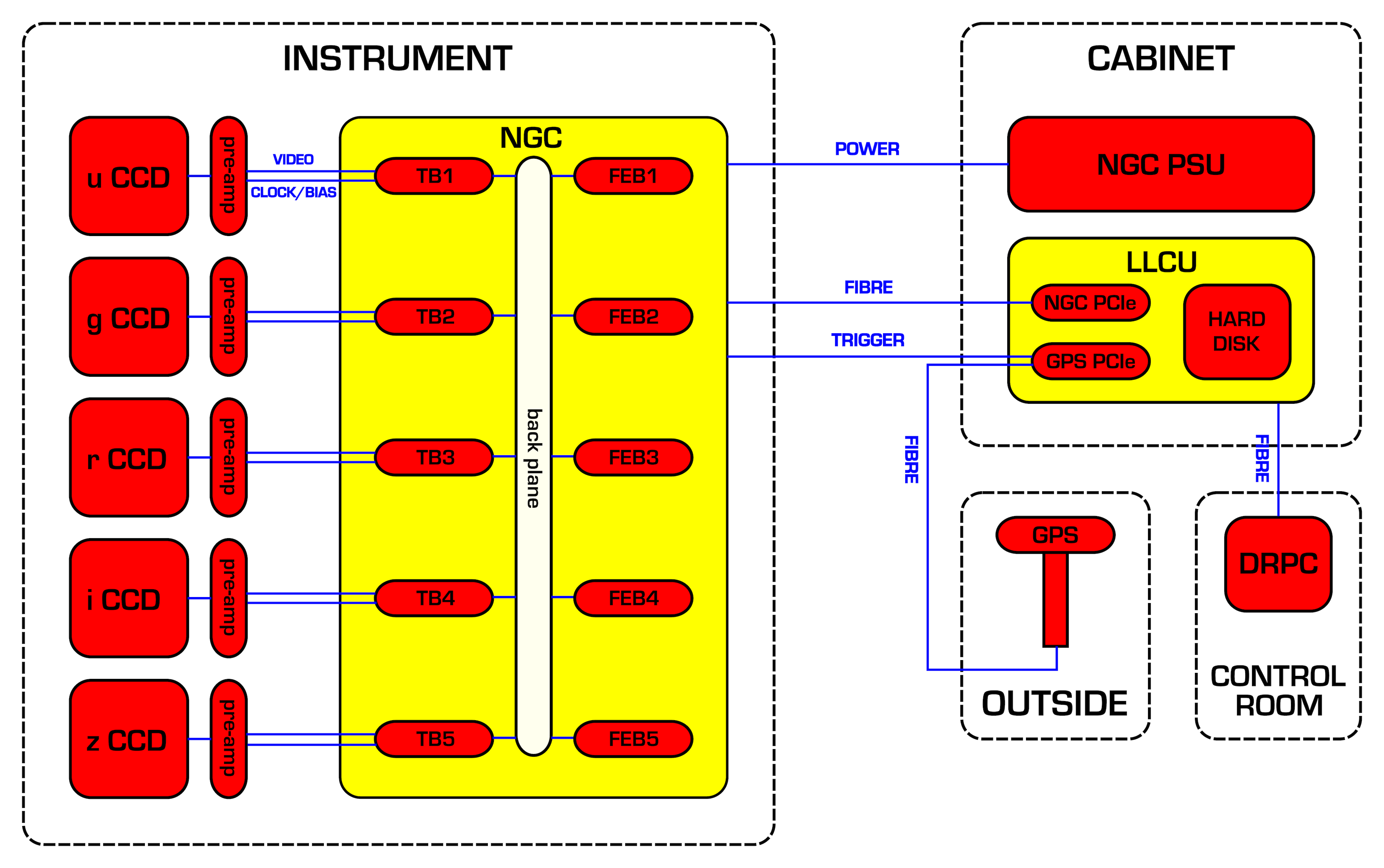}}
\caption{A block diagram showing the hardware of the HiPERCAM data
  acquisition system -- see Section~\ref{sec:hw} for details.}
\label{fig:das}
\end{figure*}

The HiPERCAM data acquisition system was designed to be detector
limited, so that the throughput of data between the CCD outputs and
the hard disk on which it is stored is always greater than the rate at
which the data can be clocked off the CCDs. This means that HiPERCAM
never has to operate in bursts, periodically pausing so that the data
archiving can catch up; instead, HiPERCAM can operate continuously all
night, even at its maximum data rate.

\subsubsection{Hardware}
\label{sec:hw}

A block diagram of the HiPERCAM data acquisition hardware is
shown in Figure~\ref{fig:das}. The architecture is similar to that
developed for ULTRACAM \citep{2007MNRAS.378..825D}, but uses much
faster and more modern hardware. At the centre of the system is a
European Southern Observatory (ESO) New General detector Controller
(NGC; \citealt{2009Msngr.136...20B}). The NGC used in HiPERCAM is
composed of a five-slot housing, with 5 Transition Boards (TB) and 5
Front-End Basic (FEB) Boards, connected via a back-plane. Each TB
handles all of the external connections to its corresponding FEB, and
is connected to a CCD via a pre-amplifier board mounted on the back of
the CCD head. The pre-amplifier board contains AC-coupled differential
pre-amplifier circuits and passive filters, and provides over-voltage
and electrostatic discharge protection on the input bias lines
\citep{2018SPIE10709E..24B}. The NGC and pre-amplifier board are
connected by a single cable that carries both the CCD video signal to
the four differential video processing chains on the associated FEB
and the clocks/biases from the clock/bias-driver on the FEB. To
minimise the length of the cables ($\sim$1.5\,m) running to each CCD
head, the NGC is located on the instrument (see
Figure~\ref{fig:mech}). In order maximise readout speed, the HiPERCAM
NGC has been configured electronically to use the Analogue Clamp
Sample (ACS) method, which takes only one sample of the voltage at the
CCD output per pixel readout cycle. The NGC can also be configured to
use slower Dual-Slope Integration (DSI), which takes two samples per
pixel, but we found this did not significantly reduce the read noise
(see Section~\ref{sec:rno}).

The NGC is powered by a separate Power Supply Unit (PSU), located in
an electronics cabinet mounted on the telescope approximately 3\,m
from HiPERCAM. The cabinet also contains a linux PC known as the LLCU
(Linux Local Control Unit). The LLCU was provided by ESO to control
the NGC, and contains the NGC Peripheral Computer Interconnect Express
(PCIe) card. The NGC PCIe card and NGC are connected by duplex fibre,
over which one can receive CCD data and control the NGC. The LLCU also
contains a large-capacity hard disk (HD) on which the raw CCD data are
written.

The LLCU contains a GPS (Global Positioning System) PCIe card made by
Spectracom (model TSync-PCIe-012) that accepts two inputs. The first
is a trigger generated by the NGC when an exposure starts, causing the
GPS card to write a timestamp to its FIFO (First In, First Out) 
buffer, which is subsequently written to the corresponding CCD frame
header. The second input is a GPS signal from an antenna located
outside the dome. The antenna and GPS card are connected by a long
(150\,m) optical fibre that isolates the telescope from lightning
strikes.

The LLCU is connected via fibre ethernet to a second linux PC located
in the telescope control room (see Figure~\ref{fig:das}), referred to
as the Data Reduction PC (DRPC). The DRPC runs the GUI (graphical user
interface) to control the instrument, the data reduction pipeline, the
target acquisition tool and the data logger, amongst other things.

\subsubsection{Software}
\label{sec:sw}

\begin{figure*}
\centerline{\includegraphics[width=13.0cm]{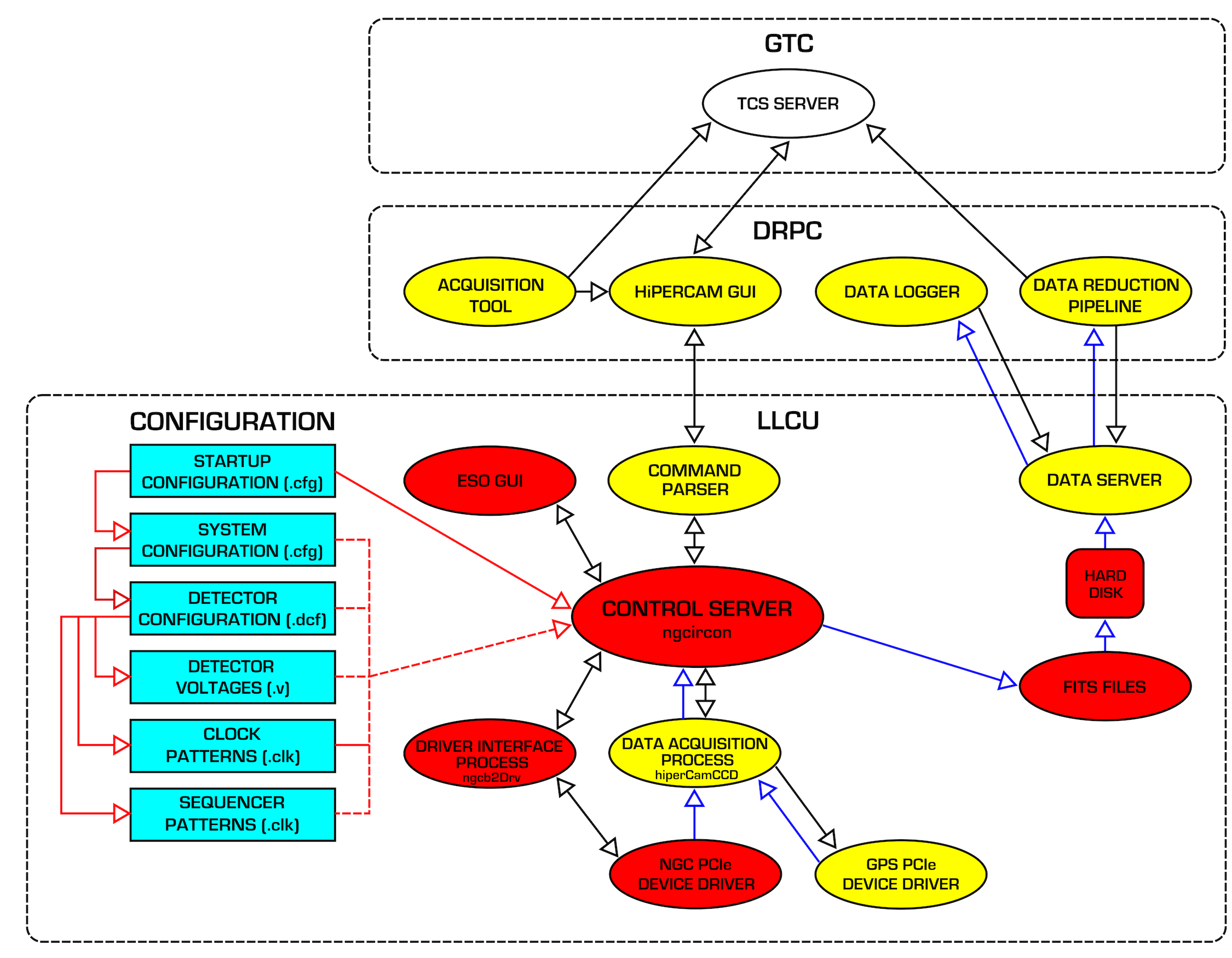}}
\caption{Software architecture and configuration of the HiPERCAM data
  acquisition system. Black arrows show the flow of commands/replies,
  blue arrows show the flow of CCD data. Red ellipses indicate tasks
  supplied by ESO as part of the standard NGCIRSW software
  distribution, yellow ellipses indicate tasks that were
  modified/written specifically for HiPERCAM, and the white ellipse
  indicates that the task was provided by the GTC.  NGC configuration
  is shown on the left in the blue boxes: solid red arrows indicate
  “specifies”, dashed red arrows indicate “reads”.}
\label{fig:sw}
\end{figure*}

The NGC is controlled using ESO's NGC Infrared Detector Control
Software (NGCIRSW), to which HiPERCAM-specific components have been
added, as shown in Figure~\ref{fig:sw}. For clarity, the first time
each of the tasks shown in Figure~\ref{fig:sw} is referred to in the
text below, it is written in italics.

Communication with the NGC is handled by the {\em Control
  Server}, which runs on the LLCU and interacts with the {\em NGC PCIe
  Device Driver} via a {\em Driver Interface Process}. The control
server can also be set up to run in simulation mode for development
and testing when no NGC is connected. The {\em Data
  Acquisition Process} (or acquisition task) is HiPERCAM specific and
also runs on the LLCU. This task begins when the ``START'' exposure
command is executed, and receives data from the CCDs via the NGC-PCIe
card. On completion of an exposure, the acquisition task reads the
timestamp from the {\em GPS PCIe Device Driver}, adds the timestamp to
the frame, and passes the data and headers to the {\em FITS Files}
task (via the control server) for writing to the hard disk. The GPS
timestamp is synchronised with the start of the exposure using an
external trigger from the NGC. The acquisition task runs
continuously until either the required number of CCD exposures have
been taken or a ``STOP'' exposure command has been issued. The
acquisition process can also perform any data pre-processing prior to
writing the frame, such as averaging multiple pixel reads for noise
reduction (see Section~\ref{sec:rno}). 

The NGCIRSW suite offers an {\em ESO GUI} (or ``engineering'' GUI), which
is useful for testing and development purposes, but for science use
has been replaced by the {\em HiPERCAM GUI} to control the NGC whilst
observing. The HiPERCAM GUI is written in Python/Tkinter and runs on
the DRPC in the control room. It communicates with the NGC on the
telescope using HTTP (Hypertext Transfer Protocol) over TCP/IP
(Transmission Control Protocol/Internet Protocol) on a dedicated
fibre-ethernet link. The interface between the HiPERCAM GUI and the
NGC is the {\em Command Parser}, which is a Python-based HTTP server
running on the LLCU with a RESTful (Representational State Transfer)
interface. The command parser translates the HTTP commands issued by
the HiPERCAM GUI to low-level NGCIRSW commands to be executed by the
NGC, e.g. to start/stop an exposure, change the CCD readout mode, or
request information on the current exposure.

The NGC configuration is set using short FITS (Flexible Image
Transport System) format files, which are editable by hand if
required. There are three types of configuration file –- startup,
system and detector configuration, as shown in
Figure~\ref{fig:sw}. The startup configuration file defines the
command-list of the control server. The system configuration defines
the NGC hardware architecture, such as the number and addresses of the
boards in the controller and LLCU. The detector configuration
describes which clock patterns, voltages and sequencer programs to
load for the setup requested by the user on the HiPERCAM GUI. The
detector voltages are defined in a voltage configuration file, in
short FITS format. The clock patterns are described in blocks, with
each block defining a sequence of clock states. Clock pattern blocks
can be defined in either hand-editable or binary format, the latter
output by the ESO graphical editing tool {\em BlueWave}. The sequencer
program defines the order of execution of the defined clock pattern
blocks and is written in Tcl/Tk.

To prepare for observing with HiPERCAM (``Phase II''), astronomers use
the {\em Acquisition
  Tool}\,\footnote{https://hcam-finder.readthedocs.io/en/latest} to
generate finding charts, specify the telescope pointing and instrument
setup, and provide cadence and signal-to-noise ratio estimates. The
required telescope pointings and instrument setups are written to JSON
(JavaScript Object Notation) files, which are also editable by
hand if required. Copies of these files are sent to the GTC Telescope Control
System (TCS), to point the telescope at the required fields, and to
the HiPERCAM GUI, to set the CCDs up for the required observations.

The HiPERCAM GUI communicates with the telescope via the {\em TCS
  Server}. This link provides information on the telescope pointing
and focus that can then be written to the FITS headers of the CCD data
files. The link also provides a way of tweaking the right
ascension, declination, rotator angle and focus of the telescope,
which is useful when acquiring targets and dithering. For the latter,
astronomers set up their required patterns using the acquisition
tool. The HiPERCAM GUI then executes the dithering pattern,
synchronising the NGC readout so that no exposures are taken whilst
the telescope is moving/settling.

\subsection{Pipeline data reduction system}
\label{sec:pipeline}

\begin{figure*}
  \centerline{\includegraphics[width=17.8cm]{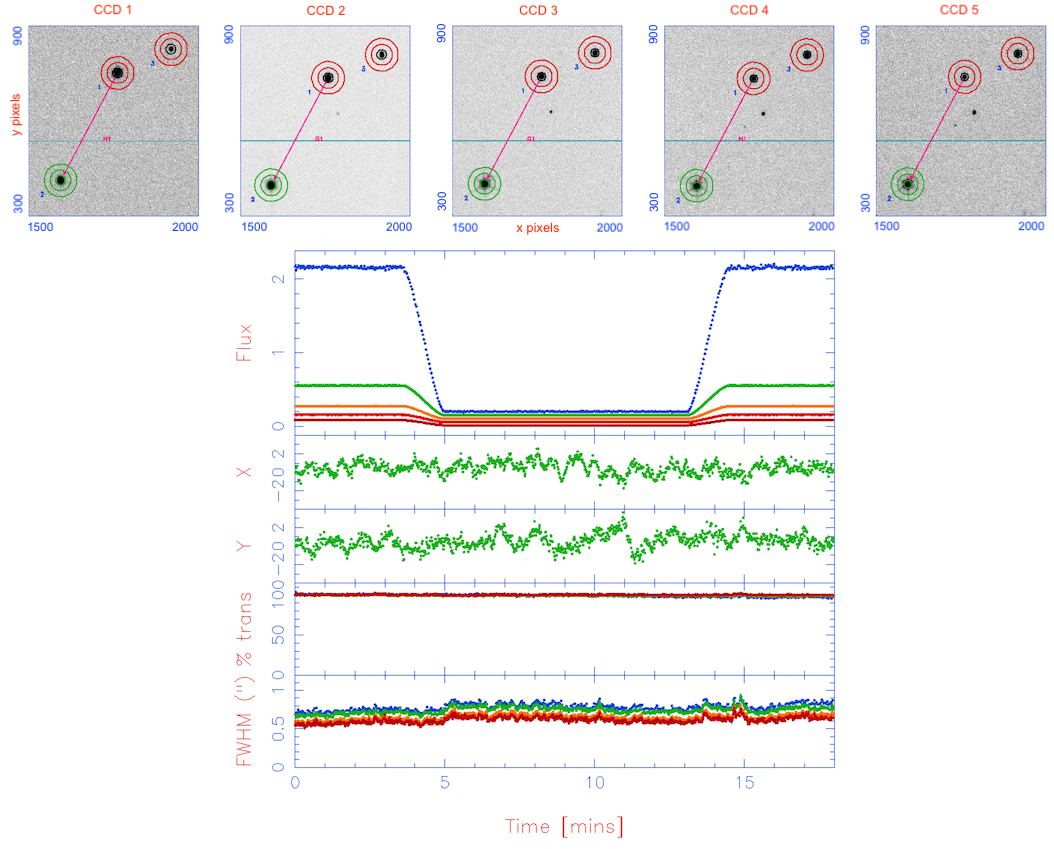}}
\caption{A screenshot of the HiPERCAM data-reduction pipeline. Top
  panel: Zoomed-in images in $u_sg_sr_si_sz_s$ of the target (the
  eclipsing red-dwarf/white-dwarf binary NN Ser) and two comparison
  stars, surrounded by software apertures defining the object and sky
  regions. Aperture 1 is the target, and apertures 2 and 3 are the
  comparison stars. Aperture 2 is green because it has been defined as
  the reference star for centroiding. The pink arrows show that the
  target and reference apertures are linked, so that the positional
  offset between the two is held constant when the target almost
  disappears during eclipse. Bottom panel: From top to bottom, the
  target flux divided by the comparison star flux in $u_sg_sr_si_sz_s$
  (each with a flux offset of 0.05 added to separate the light curves
  during eclipse minimum), the comparison star $x$ and $y$ positions,
  the sky transparency measured from the comparison star flux, and the
  seeing measured from the comparison-star FWHM in $u_sg_sr_si_sz_s$.}
\label{fig:pipeline}
\end{figure*}

HiPERCAM can generate up to 17\,MB per second of data, or up to
600\,GB per night. To cope with this relatively high data rate,
HiPERCAM has a dedicated {\em Data Reduction
  Pipeline}\footnote{http://deneb.astro.warwick.ac.uk/phsaap/hipercam/docs/html},
as shown in Figure~\ref{fig:pipeline}. The pipeline runs on the DRPC
and is written in Python. A Python TCP/IP WebSocket {\em Data Server}
running on the LLCU allows the data on the hard disk to be accessed by
the pipeline over a dedicated fibre-ethernet link (see
Figures~\ref{fig:das} and \ref{fig:sw}).  The HiPERCAM {\em Data
  Logger} accesses the same server to provide observers with a
real-time log of the data obtained.

When observing, the HiPERCAM pipeline provides a quick-look data
reduction facility, able to display images and light curves in real
time, even when running at the highest frame and data rates. Post
observing, the pipeline acts as a multi-platform, feature-rich
photometric reduction package, including optimal extraction
\citep{1998MNRAS.296..339N}. For quick-look reduction, most of the
pipeline parameters are kept hidden and the observer can easily skip
over the few that remain to view images and light curves as quickly as
possible. Conversely, when reducing data for publication, the
signal-to-noise ratio can be maximised by carefully tweaking every
parameter. The pipeline also offers an API (Application Programmers
Interface), giving users access to raw HiPERCAM data using their own
Python scripts.

To ensure efficient writing speed and storage, the raw data and
headers from a run on a target with HiPERCAM are stored in a single,
custom-format binary FITS cube. Each slice of the cube contains five
images, one from each of the HiPERCAM CCDs, and a timestamp. The file
may contain millions of such slices if a high-speed observation is
performed. The pipeline grabs these individual slices, or frames, for
processing, and can write out standard-format FITS files containing a
single exposure from the five HiPERCAM CCDs, if required.

Although autoguiding is provided at the GTC Folded Cassegrain focus
used by HiPERCAM, secondary guiding from the science images is useful
in cases where no guide stars can be found or to eliminate the effects
of flexure between HiPERCAM and the guide-probe arm. Using stellar
centroids calculated as part of the real-time data reduction, the
pipeline is able to send regular right ascension and declination
offsets to the TCS server to correct for any tracking errors.
 
\subsection{Readout modes}
\label{sec:modes}

HiPERCAM can be read out in three different modes: full-frame,
windowed and drift mode, as shown in Figure~\ref{fig:ccd}. In full-frame
mode, the entire image area is read out, with an option to include the
over-scan and pre-scan regions for bias-level determination. The
windowed mode offers either one window in each quadrant (or one
``quad'') or two windows in each quadrant (two quads), with an option
to include the pre-scan (but not over-scan) regions. Drift mode is for
the highest frame-rate applications, and uses just two windows lying
at the border between the lower image and storage areas, as shown in
Figure~\ref{fig:ccd} and described in greater detail below.

To ensure that the five HiPERCAM CCDs read out simultaneously, and to
simplify the data acquisition system, each window in a quad must have
the same pixel positions in all five detectors. In addition, the data
acquisition system expects data from each of the four outputs of the
CCDs to be processed at the same time, which effectively means that
the windows in a quad must lie the same number of pixels from the
vertical centre-line of the detector. This restriction would make
target acquisition difficult, so in practice a {\em differential
  shift} is performed during readout: The three windows in a quad
lying furthest from their respective CCD outputs are shifted along the
serial register to lie at the same distance from the output as the
closest (fourth) window of the quad. A detailed description of the
differential shift technique is given in Appendix~A2 of
\citet{2007MNRAS.378..825D}.

The only restrictions on the sizes and positions of the windows are
that they must not overlap with each other or with the borders between
the readout quadrants, and that the windows in each quad must have
identical sizes and vertical start positions. All windows must also
have the same binning factors; pixels can be binned by factors of 1 to
12 in each dimension. These restrictions simplify the data acquisition
system but still give flexibility in acquiring targets and
comparison stars in the windows by adjusting the horizontal/vertical
positions and sizes of the windows, the telescope position, and the
instrument rotator angle. The HiPERCAM acquisition tool (see
Section~\ref{sec:sw}) can be used to assist in this process.

The CCDs in HiPERCAM are split frame-transfer devices, as shown in
Figure~\ref{fig:ccd}. When an exposure is finished, each image area is
shifted into its corresponding storage area, and the next exposure
begins. This frame-transfer process is quick -- 7.8\,ms with slow
clocking (the corresponding figure for the fast clock setting is
6.7\,ms). During an exposure, the previous image in the storage area
is vertically shifted into the serial register row-by-row, with any
unwanted rows between the windows being dumped. Each row is then
horizontally-clocked along the serial register to the output where it
is digitised\footnote{The word {\em digitisation} here refers to both
  the determination of the pixel charge content via ACS and the
  digitisation of the signal by the ADC. The frequency at which this
  occurs is referred to as the {\em pixel rate.}}, with any unwanted
pixels in the serial register lying either side of the windows being
dumped. Therefore, whilst the current frame is exposing, the previous
frame is being read out. The dead time between exposures is thus
only 7.8\,ms in HiPERCAM -- the time it takes to shift the image into
the storage area. The rapid shifting from image to storage area acts
like an electronic shutter, and is much faster than a conventional
mechanical shutter. The lack of mechanical shutters in HiPERCAM does
result in vertical trails in short-exposure images of bright stars,
but these can be overcome in some situations using the focal-plane
mask (see Section~\ref{sec:mech}).

As well as two different clocking speeds (slow/fast), HiPERCAM also
offers two pixel rates (slow/fast), as indicated in
Figure~\ref{fig:ccd}. Using the slow clock and pixel speeds, a full
frame can be read out every 2.9\,s with a dead time of 7.8\,ms; the
corresponding figures for the fast clock and pixel speeds are 1.1\,s
and 6.7\,ms, respectively.

It is more complicated to set a precise exposure time with HiPERCAM
than it is with a non-frame-transfer CCD. This is because it is not
possible to shift the image area into the storage area until there is
sufficient room in the storage area to do so. The fastest exposure
time is therefore given by the time it takes to clear enough space in
the storage area, which in turn depends on the window sizes, locations
and binning factors, as well as the clocking and pixel rates, all of
which are variables in the HiPERCAM data acquisition system. If an
exposure time longer than the time it takes to read out the storage
area is required, an {\em exposure delay} must be added prior to the
frame transfer to allow photons to accumulate in the image area for
the required amount of time. On the other hand, if a shorter exposure
than the time it takes to read out the storage area is required, the
exposure delay must be set to zero and the binning, window and
clocking/pixel rates adjusted so that the detector can frame at the
required rate. Since the frame transfer time, i.e. the time required
to vertically clock the whole image area into the storage area, is
7.8\,ms in HiPERCAM, the maximum frame rate is limited to
$\sim$122\,Hz, but with a duty cycle (the exposure time divided by the
sum of the exposure and dead times) of less than 5 per cent. With a
more useful duty cycle of 75 per cent, the maximum frame rate is only
$\sim$30\,Hz.

For frame rates significantly faster than $\sim$30\,Hz, a different
readout method is required, known as {\em drift mode}. We originally
developed this mode for ULTRACAM and ULTRASPEC (see
\citealt{2007MNRAS.378..825D}, \citeyear{2014MNRAS.444.4009D}), and
the readout sequence is shown pictorially and described in detail in
Figure~A1 and Appendix~A of \citet{2007MNRAS.378..825D}. Two windows,
one for the science target and the other for a comparison star, are
positioned at the bottom of the image area, next to the border with
the storage area (see Figure~\ref{fig:ccd}). At the end of an
exposure, only the two windows, not the entire image area, are
vertically clocked into the (top of) the storage area. The results in
a stack of windows being present in the storage area at any one time,
and a dramatic reduction in the dead time between exposures because it
is now limited by the time it takes to move a small window rather than
the full frame into the storage area. For example, two 4$\times$4
binned HiPERCAM windows of size 32$\times$32 pixels would take only
0.4\,ms to move into the storage area in drift mode, providing a frame
rate of $\sim$600\,Hz with a duty cycle of 75 per cent -- a factor of
20 improvement over windowed mode\footnote{A HiPERCAM frame-rate
  calculator can be found at: \newline
  http://www.vikdhillon.staff.shef.ac.uk/hipercam/speed.html}.

Due to its complexity, drift mode only allows two windows to be used,
with no pre-scan or over-scan regions and no clearing between frames.
The only difference in how drift mode has been implemented in
HiPERCAM compared to ULTRACAM is that two additional windows are read
out by the upper two outputs of the HiPERCAM CCDs during drift mode,
but the top half of the image area is obscured by the focal-plane mask
and hence these windows are not processed by the data reduction
pipeline.

Although drift mode has a clear advantage in terms of frame rate, it
has the disadvantage that only two windows, instead of up to eight,
are available. Also, drift mode windows spend more time on the CCDs,
and hence accumulate more sky photons and more dark current. Hence,
although the additional sky photons can be blocked by the focal-plane
mask, and the dark current in HiPERCAM is negligible, it is
recommended that drift mode should only be used when the duty cycle in
non-drift mode becomes unacceptable, which typically occurs when frame
rates in excess of about 30\,Hz are required.

When observing bright standard stars or flat fields it is sometimes
necessary to take full-frame images with short exposure times. HiPERCAM
therefore offers users the option of taking exposures of arbitrarily
short length by turning CCD clearing on. When clearing is on, data in
the image area are dumped prior to exposing for the required length of
time. Hence any photo-electrons collected in the image area whilst the
previous exposure is reading out are discarded. Clear mode has the
disadvantage that the duty cycle is poor (25 per cent in the case of a
full-frame 1\,s exposure).

By default, all HiPERCAM CCDs start and end their exposures at exactly
the same time. This synchronicity is of great scientific value when
comparing the variability of sources at different wavelengths, but can
result in significant signal-to-noise ratio variations between the
bands if an object is particularly blue or red. It is possible for
each of the HiPERCAM CCDs to have a different exposure time, and still
ensure strict simultaneity of readout, by skipping the readout of
selected CCDs using the {\em NSKIP} parameter in full-frame and
windowed mode. For example, setting the exposure time to 2\,s and
NSKIP to 3,2,1,2,3 for the $u_s,g_s,r_s,i_s,z_s$ CCDs would result in
the NGC reading out only the $r_s$-band CCD on the first readout cycle
(giving a 2\,s $r_s$ exposure), then the $g_s$, $r_s$ and $i_s$-band
CCDs on the second readout cycle (giving a 4\,s $g_s$ and $i_s$
exposure and a 2\,s $r_s$ exposure), and then the $u_s$, $r_s$ and
$z_s$-band CCDs on the third readout cycle (giving a 6\,s $u_s$ and
$z_s$ exposure and a 2\,s $r_s$ exposure), etc.

The $u_s$, $r_s$ and $i_s$-band images experience an odd number of
dichroic reflections, as shown in Figure~\ref{fig:raytrace}b, and must
therefore be corrected for the left-right flip compared to the $g_s$
and $z_s$-band images.  This is achieved by swapping the serial
clocking between the E and H outputs, and the F and G outputs, in the
$u_s$, $r_s$ and $i_s$ CCDs (see Figure~\ref{fig:ccd}). It is possible
to swap the outputs in this way on individual CCDs because the
sequencer scripts (see Figure~\ref{fig:sw}) for each CCD run on
separate FEBs (see Figure~\ref{fig:das}). An alternative option would
have been to perform this output swapping by altering the cables
between the $u_sr_si_s$ CCDs and the NGC, but it is preferable from a
cable design, manufacture and maintenance perspective to have
identical cables for all CCDs. Note that swapping the serial clocking
is only necessary in windowed and drift modes -- it is not required
for full-frame readout as the image flip can be corrected in the data
reduction pipeline.

\section{Performance on the GTC}
\label{sec:performance}

HiPERCAM saw first light on the GTC in February 2018, and it has since
been in use for 13 observing runs totalling $\sim$70 nights (although
some of these nights were shared with other instruments or partly lost
due to weather). The first tranche of scientific papers based on
HiPERCAM data have now been published (\citealt{2019NatAs...3..553R},
\citealt{2019ApJ...883...42N}, \citealt{2019MNRAS.490L..62P},
\citealt{2020NatAs...4..690P}, \citealt{2020ApJ...891...45K},
\citealt{2021arXiv210511769P}, \citealt{2020ApJ...905...32B},
\citealt{2020ApJ...898L..25K}, \citealt{2021arXiv210610283M},
\citealt{2021arXiv210707573V}). In this section, we summarise the
performance of HiPERCAM on the GTC.

\subsection{Image quality}

We measured the plate scale of each CCD by stepping a bright star
across the field of view by a known angle and measuring the movement
in pixels. We find the same value of the plate scale in all five bands
to within the errors, $0.0805\pm 0.0001$\,arcsec/pixel, as
required (see Section~\ref{sec:reqts}).

In order to assess the image quality, we observed a dense stellar
field during excellent seeing conditions, after aligning and focusing
the secondary and segmented primary mirrors. The FWHM of stars at the
centre of the images in each filter were measured to be
0.56/0.44/0.41/0.37/0.36 ($\pm$0.02) arcsec in $u_sg_sr_si_sz_s$,
respectively, with no significant deviations from these values in the
corners of the field of view, as required (see
Section~\ref{sec:reqts}). HiPERCAM on the GTC can therefore provide
images that are seeing-limited across the whole field of view in even
the best observing conditions on La Palma.

We do not expect to see any ghosting in HiPERCAM images (see
Section~\ref{sec:optics}) because the dichroics operate in a
collimated beam and have anti-reflection coatings on their rear
surfaces (see Section~\ref{sec:optics}). This is indeed the case --
the brightest (saturated) stars in the images show no discernible
ghosting, down to a level given by the read noise, i.e. less than one
part in 10$^4$. The pixel positions of the stars are the same to
within approximately 5 pixels (75\,$\mu$m) on all five CCDs, showing
that the relative alignment of the CCD heads is good. We measured the
vignetting from images of blank regions of the twilight sky, finding
the field of view to be flat from the centre to the corners to better
than $\sim$5 per cent.

\subsection{Throughput and sensitivity}

The HiPERCAM zero points on the GTC, defined as the magnitude of a
star above the atmosphere that gives 1 photo-electron per second in
each filter, were measured from SDSS standard star observations
\citep{2002AJ....123.2121S} on photometric, non-dusty nights. We found
values of 28.15/29.22/28.78/28.43/27.94 in $u_sg_sr_si_sz$,
respectively. The errors on these zero points are estimated to be
$\pm0.05$ and are dominated by the uncertainty in the primary
extinction coefficients, which we measured from the light curves of
comparison stars observed on the same nights as the standards; we
found typical atmospheric extinction values of
0.48/0.17/0.10/0.05/0.05 in $u_sg_sr_si_sz$, respectively, consistent
with the long-term values measured at the Observatorio del Roque de
los
Muchachos\footnote{https://research.ast.cam.ac.uk/cmt/camc\_extinction.html}.
For comparison, OSIRIS \citep{2003SPIE.4841.1739C}, the workhorse
single-beam, red-optimised imaging spectrograph on the GTC, has
observed zero points of 25.76/28.26/28.84/28.49/27.95 in $ugriz$,
respectively, demonstrating that HiPERCAM is competitive with OSIRIS
in the red and superior in the blue.

\begin{figure}
\centerline{\includegraphics[width=8.6cm]{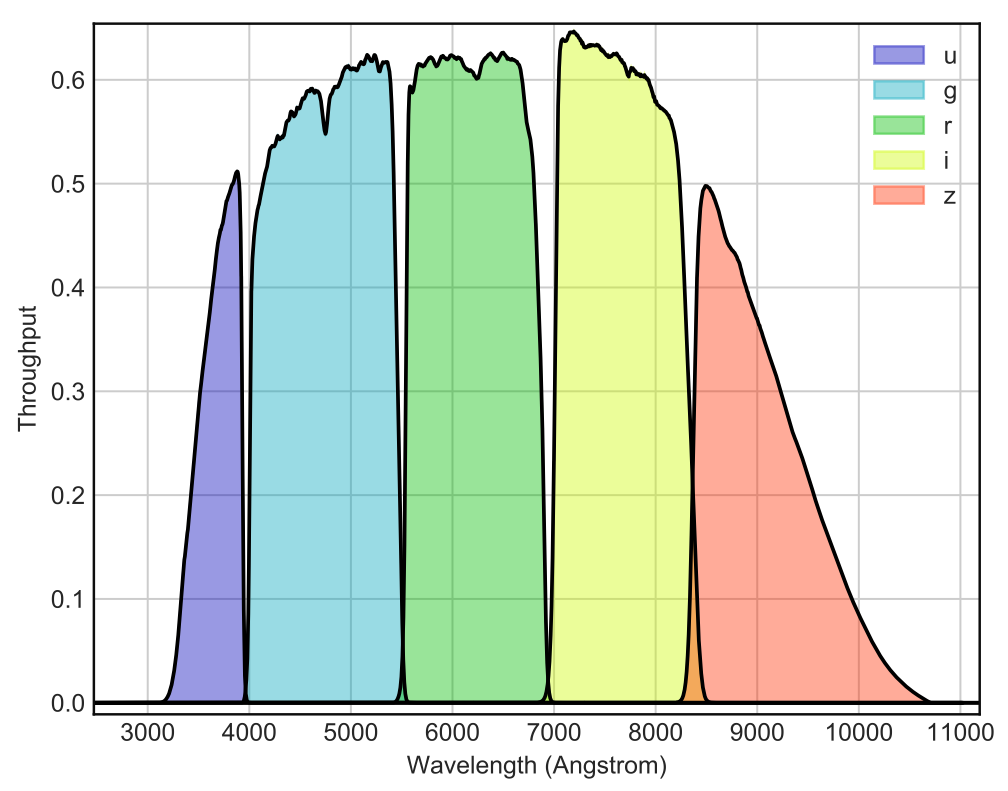}}
\caption{HiPERCAM throughput in the $u_sg_sr_si_sz_s$ bands, not
  including the telescope and atmosphere.}
\label{fig:thruput}
\end{figure}

We have estimated the response of the HiPERCAM optics and detectors by
building a throughput model based on the measured lens/window/filter
transmissions, dichroic reflectivities/transmissions and the CCD
QEs\footnote{The model uses the Python module {\em pysynphot} and is
  available from:
  https://github.com/StuartLittlefair/ucam\_thruput.}. The throughput
is shown as a function of wavelength in Figure~\ref{fig:thruput}. It
can be seen that the throughput peaks at over 60 per cent in
$g_sr_si_s$, exceeds 50 per cent in $u_s$ and $z_s$, and there is some
sensitivity even up to 1060\,nm. This is more efficient than many
single-beam imagers, e.g. the throughput of Keck/LRIS is 26/51/48 per
cent in
$BVR$\footnote{https://www2.keck.hawaii.edu/inst/lris/photometric\_zero\_\newline
  points.html}, respectively, despite the fact that HiPERCAM also has
dichroic beamsplitters in the light path. The high throughput of
HiPERCAM has been achieved by using CCDs and high-performance,
multi-layer coatings on the dichroics, filters, lenses and windows
that have each been optimised for operation in their bandpass, rather
than for all bandpasses. Using the throughput model, we calculate
theoretical HiPERCAM zero points on the GTC of
28.09/29.22/28.86/28.52/27.92 in $u_sg_sr_si_sz$, respectively, which
agree to within a few per cent with the observed zero points,
demonstrating that the instrument is performing to specification. A
detailed analysis of the HiPERCAM colour terms when using the Super
SDSS filters is deferred to another paper (Brown et al., in
preparation).

\begin{figure}
\centerline{\includegraphics[width=8.6cm]{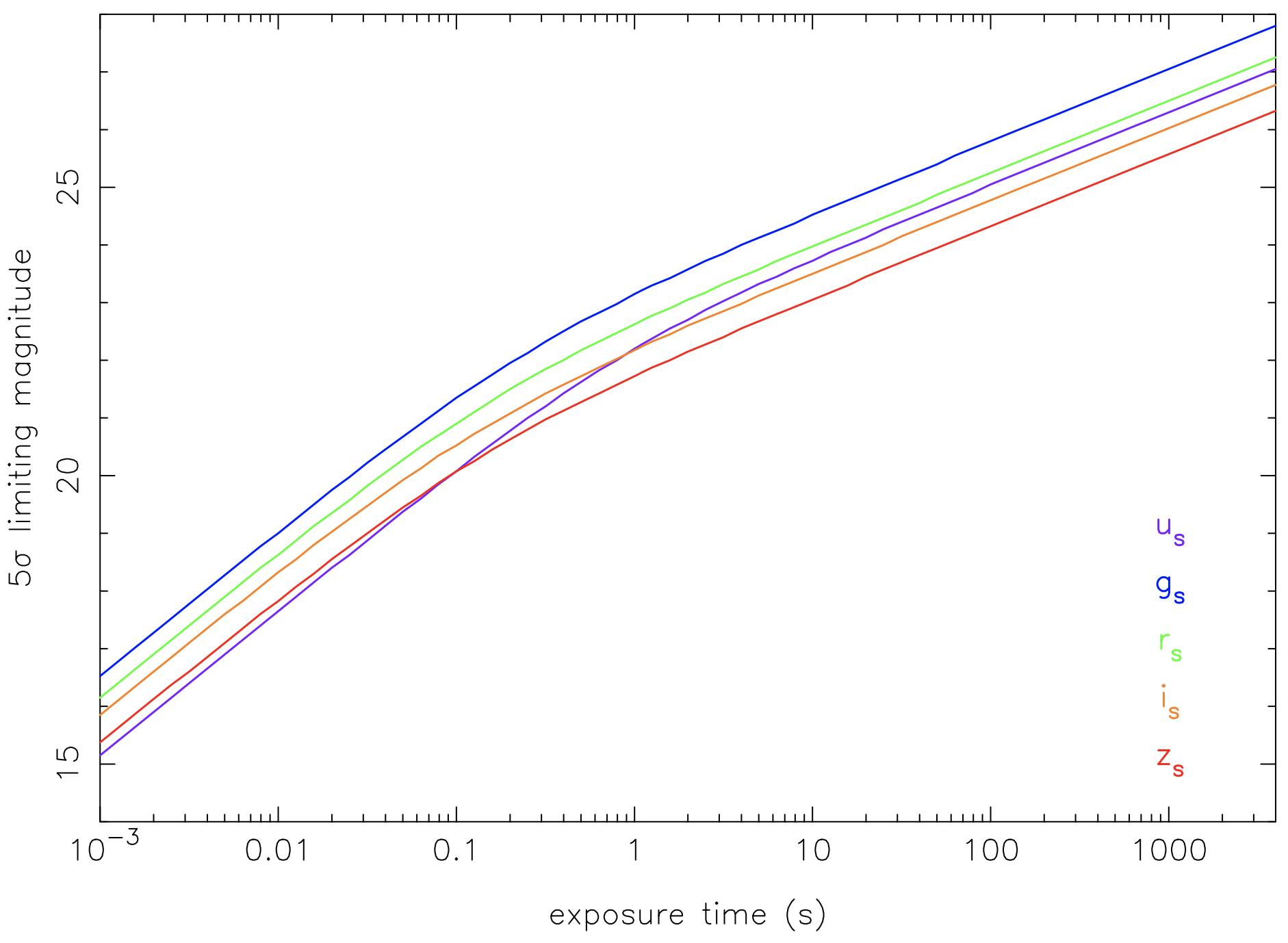}}
\caption{Limiting magnitudes (5$\sigma$) of HiPERCAM on the GTC as a
  function of exposure time in $u_s g_s r_s i_s z_s$ (purple, blue,
  green, orange and red curves, respectively), assuming seeing of
  0.6\,arcsec, dark moon and observing at the zenith.}
\label{fig:snr}
\end{figure}

Figure~\ref{fig:snr} shows the limiting magnitudes of HiPERCAM on the
GTC as a function of exposure time, calculated from the measured zero
points. It is possible to obtain 5$\sigma$ limiting magnitudes of
$g_s\sim23$ in 1\,s and $g_s\sim28$ in 1\,h\footnote{A signal-to-noise
  ratio calculator for HiPERCAM+GTC is available at: \newline
  http://www.vikdhillon.staff.shef.ac.uk/hipercam/etc.html}.

\subsection{Read noise and cross talk}
\label{sec:rno}

The read noise of HiPERCAM is limited by the bandwidth of the
pre-amplifier, which is currently hard-wired to 1.06\,MHz
\citep{2018SPIE10709E..24B} -- see Section~\ref{sec:rno_enhance} for a
future enhancement that will overcome this restriction. As a result,
HiPERCAM currently has a read noise of $\sim$5.5\,e$^-$ at the fast
pixel rate of 526\,kHz, and $\sim$4.5\,e$^-$ at the slow pixel rate of
192\,kHz (which involves averaging 4 samples of the charge content of
each pixel in the NGC, each taken at $\sim$1\,MHz). These read noise
values were measured at the GTC using the dummy outputs of the CCDs to
eliminate pickup noise, which theoretically increases the read noise
by approximately a factor of $\sqrt{2}$ compared to so-called {\em
  single-ended mode}. However, we see significant pick-up noise at the
telescope in single-ended mode and hence always use the dummy outputs
whilst observing.

We also checked for cross talk, due to the multiple outputs on the
CCDs and their associated electronics. When a bright source is present
in one of the CCD quadrants, cross talk manifests itself as a ghost
image (positive or negative) at a mirrored position in the other
quadrants \citep{2001ExA....12..147F}. We searched the mirror positions
to bright (saturated) sources in HiPERCAM images and found no evidence
for any cross-talk signal, down to a level given by the read noise,
i.e. less than one part in 10$^4$.

\subsection{Timestamping}

Given that HiPERCAM can image at rates exceeding 1\,kHz, it is
important that each frame is timestamped to a significantly better
accuracy than this, i.e. to better than $\sim$100\,$\mu$s. To measure
the timestamping accuracy, we observed an LED mounted on the
focal-plane mask with HiPERCAM. The LED was triggered by the
pulse-per-second (PPS) output of the GPS card to turn on precisely at
the start of each UTC second, and off again half a second later. The
accuracy of the PPS output is better than 50\,ns and the LED rise time
is of a similar order, so these are insignificant sources of error.
The LED formed a pseudo-star in the images, which were reduced by
the HiPERCAM data-reduction pipeline. The resulting light curves were
then phase-folded on the 1\,s period of the LED. The light curve shape
is a convolution of two top-hat functions, one for the exposure time
duration and the other for the LED pulse, resulting in the folded
light curve showing a ramp. In the absence of any timestamping errors,
the centre of the ramp should correspond to the start of the UTC
second. We tested every readout mode in this way\footnote{For details,
  see: http://deneb.astro.warwick.ac.uk/phsaap/\newline
  hipercam/docs/html/timing/timing\_tests.html} and found that the LED
turned on within $\sim$100\,$\mu$s of the start of each UTC second,
thereby meeting the absolute timestamping accuracy requirement of
HiPERCAM. This test is insensitive to timestamping errors equal to an
integer number of seconds, but it is difficult to see how such an
error could arise in the HiPERCAM data acquisition system, and we
would have noticed such a large error in our multi-instrument
monitoring of eclipsing white dwarfs,
e.g. \citet{2014MNRAS.437..475M}.

We also measured the frame-to-frame stability of the HiPERCAM exposure
times by measuring the time intervals between 5 million
consecutive HiPERCAM drift-mode observations taken with a frame rate of
1\,kHz: the exposure times remained constant to better than 100\,ns.

\subsection{Flexure}

Whilst observing a star field, we turned the rotator through
180$^{\circ}$ and determined the location of the rotator centre, which
we found lay (4,\,12) pixels from the $r_s$-band CCD centre, verifying
that the mechanical alignment of HiPERCAM is excellent. This
measurement was made near the zenith and was then repeated at an
altitude of approximately 40$^{\circ}$. We found that the rotator
centre values on all 5 CCDs were consistent between the zenith and
40$^{\circ}$ to within one pixel, indicating mechanical flexure of
less than 15\,$\mu$m at the detector, as required (see
Section~\ref{sec:mech}).

\subsection{Reliability}

HiPERCAM currently has only one moving part -- the focal-plane slide,
and hence it is an intrinsically reliable instrument. We estimate that
we have lost less than 5 per cent of observing time due to technical
problems with the instrument during the $\sim$70 nights that HiPERCAM
has been in use on the GTC to date. The majority of this time loss has
been due to problems with the flow sensors and the CCD vacuum
seals. The flow-sensor problem has now been rectified by switching
from the original Hall-effect flow sensors (which failed due to
metallic particles in the coolant clogging up the magnetic rotors), to
ultrasonic flow sensors (which also failed due to the presence of
micro-bubbles in the coolant), to optical flow sensors (which appear
to work well). The problems with the loss of vacuum in some of the CCD
heads were mostly due to the copper gaskets used for the main case
seals and have since been rectified by resealing.

\section{Future plans}

With HiPERCAM now working to specification and entering its science
exploitation phase at the GTC, we have begun a program of instrument
enhancements to further improve its performance.

\subsection{COMPO}
\label{sec:compo}

To correct for transparency variations in the Earth's atmosphere,
astronomers use the technique of differential photometry, where the
target flux is divided by the flux of one or more comparison stars
observed at the same time, and in the same patch of sky, as the
target. In order not to degrade the signal-to-noise ratio of the
resulting light curve significantly, it is necessary to use comparison
stars that are brighter than the target star. If the target star is
particularly bright, it becomes difficult to find suitable comparison
stars, especially if the field of view of the photometer is small.

The probability of finding a comparison star of a given magnitude
depends on the galactic latitude of the target and the search radius,
and can be calculated from the star counts provided by
\citet{simons95}. If the search radius is equal to the 3.1\,arcmin
(diagonal) field of view of HiPERCAM, the probability of finding a
comparison star of magnitude $r_s = 14$ is 90 per cent at a Galactic
latitude of 30$^{\circ}$ (the all-sky average)\footnote{A comparison
  star probability calculator is available at: \newline
  http://www.vikdhillon.staff.shef.ac.uk/ultracam/compstars.html}.
Such a comparison star would be fainter than the brightest scientific
targets observed with HiPERCAM, such as the host stars of transiting
exoplanets, thereby limiting the signal-to-noise ratio of the
differential light curve. In addition, most comparison stars are
likely to be red, whereas the majority of HiPERCAM targets are blue,
exacerbating the problem in the $u_s$-band in particular.

\begin{figure*}
  \centerline{\includegraphics[width=15.0cm]{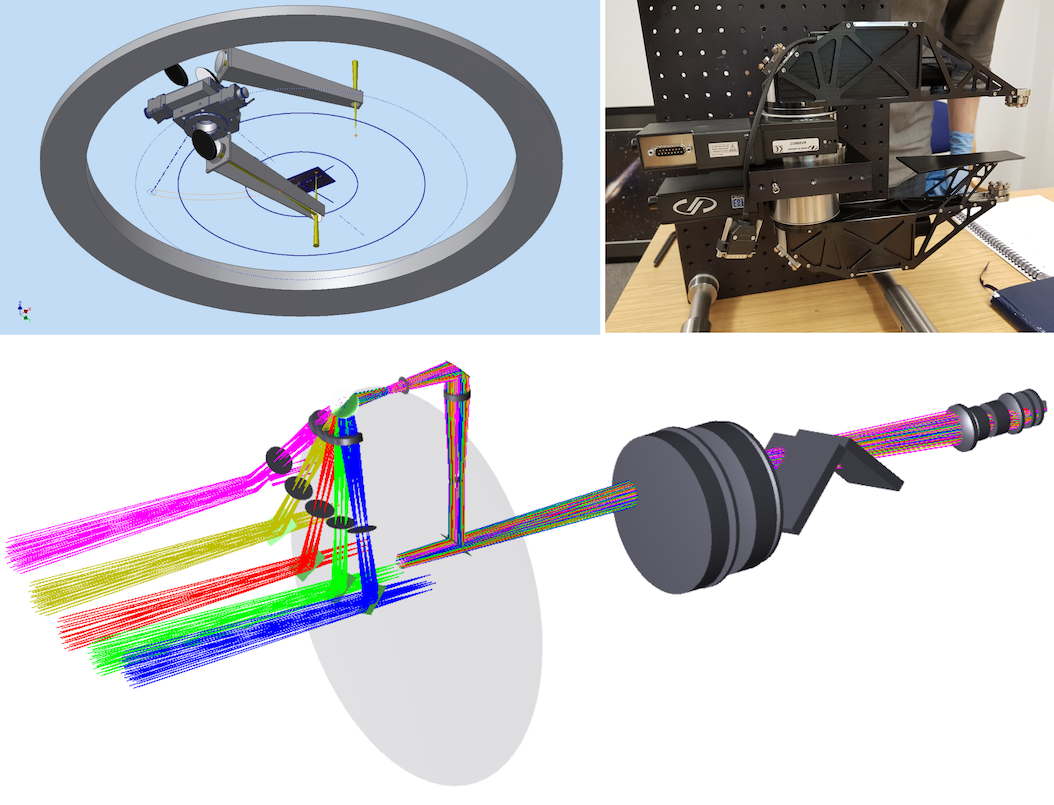}}
\caption{{\em Top left:} Schematic of COMPO, looking up at the
  telescope focal plane. The HiPERCAM field of view is indicated by
  the filled blue rectangle at the centre. The upper (pick-off) arm
  collects light from a star, indicated by the upper yellow cone of
  light, falling outside the HiPERCAM field of view but inside the
  10.3\,arcmin diameter view of view at the GTC Folded Cassegrain
  focus (outer solid blue circle). The lower (injection) arm redirects
  this light via some relay optics to one of the corners of the
  HiPERCAM field of view, indicated by the lower yellow cone of
  light. {\em Top right:} Photograph of COMPO during assembly in the
  lab, showing the pick-off arm (top) and injection arm (bottom)
  attached to their respective rotation stages. The baffle attached to
  the injection arm can be seen at the right. {\em Bottom:} Ray trace
  through the COMPO optics. Light from the GTC at the left is incident
  on the pick-off mirror, shown at five different off-axis angles by
  the coloured beams. The light first passes through a field stop and
  collimator lens in the pick-off arm, which are mounted on a
  motorised linear stage to compensate the focus for the curved
  telescope focal plane (shown by the large, light-grey ellipse). The
  light is then redirected to the injection arm via two fold mirrors,
  and passes through a re-imaging lens and another field stop before
  being reflected by the injection mirror into HiPERCAM (displayed on
  the right with dark grey lenses/dichroics -- only the $z_s$ arm is
  shown).}
\label{fig:compo}
\end{figure*}

One way to increase the brightness of comparison stars available for
differential photometry is simply to increase the field of view of the
instrument. In the case of HiPERCAM, this can be achieved by replacing
the existing collimator with a larger one, as described in
Section~\ref{sec:optics}. However, this would be extremely expensive
and would only increase the diagonal field of view to 4.3\,arcmin,
giving a 90 per cent probability of finding a comparison star of
magnitude $r_s = 13$, i.e. gaining only one magnitude in brightness.
A much more effective and cheaper solution is to use the
COMParison star Pick-Off system (COMPO) shown in
Figure~\ref{fig:compo}. Light from a bright comparison star that falls
within the 10.3\,arcmin diameter field of view of the GTC Folded
Cassegrain, but outside the 3.1\,arcmin field of view
of HiPERCAM, is collected by a pick-off arm lying just above the
telescope focal plane. The light is then redirected to a second arm
lying just below the focal plane, via some relay optics, which
injects the starlight onto one of the bottom corners of the HiPERCAM
CCDs. The effective field of view for comparison stars is hence 
increased to 6.7\,arcmin, giving a 90 per cent probability of finding
a comparison star of magnitude $r_s = 12$, i.e. gaining two magnitudes
in brightness.

The pick-off and injection arms rotate around a fixed point that lies
outside the patrol field, as shown in Figure~\ref{fig:compo}. The
rotation axis of the arms is tilted to align it with the telescope
exit pupil. The internal relay optics of the pick-off arm include a
collimator and field stop to eliminate off-axis rays and help control
stray light, which are together mounted on a motorised linear stage
that moves along the optical axis to compensate for the telescope
focal-plane curvature.

Attenuation of the comparison-star light within COMPO is unimportant,
as long as it remains constant during the observation. The pick-off
has a square field of view of side 24\,arcsec and this is
injected onto a square of side 330 pixels in the corner of each CCD. A
baffle mounted at the end of the injection arm and positioned close to
the telescope focal plane is used to prevent any light scattered by
the COMPO arms from entering the instrument and also gives a sharp
edge to the injected field in the final image. The rest of the field
of view of HiPERCAM is unaffected by COMPO, so any other comparison
stars that fall in the image can be used as before. Users will be able
to select suitable comparison stars for COMPO using the acquisition
tool described in Section~\ref{sec:sw}. For users who do not
need to use COMPO, the arms can be fully retracted out of the beam.

The manufacture of COMPO is now complete and we hope to commission the
system at the GTC during 2022.

\subsection{Diffuser}

When observing the brightest sources with HiPERCAM, such as the host
stars of transiting exoplanets, the signal-to-noise ratio in a
differential light curve can be limited by variations in seeing or
atmospheric scintillation, rather than the brightness of the target or
comparison stars (see \citealt{2015MNRAS.452.1707O} and
\citealt{2019MNRAS.489.5098F}). In the case of seeing, the varying PSF
alters the fraction of light falling outside the photometry software
aperture in a way that differs between the target and comparison
stars, due to the fact that the latter almost always lie outside the
isoplanatic patch (only $\sim$2\ arcsec in the optical on La Palma;
\citealt{1994A&A...284..311V}) of the former. Simply increasing the
size of the software aperture is not a solution due to the
corresponding increase in sky and read noise, and profile fitting is
unable to model the subtle changes in PSF due to rapid seeing
variations. It is possible to create a more stable PSF by defocusing
the telescope (e.g. \citealt{2009MNRAS.396.1023S}), but the most
stable PSFs are only achievable using beam-shaping diffusers
\citep{2017ApJ...848....9S}.

We have tested such a diffuser in HiPERCAM on the GTC. The diffuser
was placed in front of the collimator and, as expected, gave a much
more stable PSF compared to using telescope defocusing. The diffuser
we tested was not optimised for HiPERCAM -- the diameter of the
diffuser was only 150\,mm, rather than the required 225\,mm, and hence
there was vignetting at the edge of the HiPERCAM field of view. In
addition, the throughput of the diffuser fell from $>$90 per cent in
$g_sr_si_sz_s$ to $\sim$70 per cent in $u_s$, due to the non-optimised
diffuser polymer, substrate and AR coatings used.  Therefore, it is
our intention to procure a new, custom diffuser for HiPERCAM that has
a larger diameter and higher $u_s$-band throughput. By combining this
new diffuser with COMPO, HiPERCAM on the GTC will become the perfect
tool for ground-based, broadband transmission-photometry studies of
the atmospheres of transiting exoplanets.

\subsection{Read noise}
\label{sec:rno_enhance}

We aim to reduce the read noise of HiPERCAM to approximately 3\,e$^-$
using a combination of measures, including: introducing a
software-switchable bandwidth in the pre-amplifier so that reduced
bandwidths in combination with slower pixel rates can be used to reach
the read-noise floor of the system; reducing the bandwidth
in the NGC FEBs from the current value of 3.9\,MHz to approximately
2\,MHz; reducing the voltage noise of the op-amps and the resistance
of the resistors (to reduce their thermal noise) in the pre-amplifier;
replacing the bias and clock cables running between the pre-amplifier
and NGC with twisted pairs and braided shields. All of these
modifications are now under test in the lab, and the most effective
ones will be implemented in HiPERCAM during 2022.

\subsection{New rotator}

HiPERCAM is currently mounted on the Folded Cassegrain E focus of the
GTC, which it currently shares with at least two other
instruments\footnote{http://www.gtc.iac.es/instruments/instrumentation.php}.
This means that HiPERCAM has to be mounted/dismounted from the
telescope once or twice a year, restricting the amount of available
telescope time and the fraction of sky that can be accessed, and results
in HiPERCAM sometimes being unavailable for following-up exciting
new astronomical transients. Sharing the focus with other instruments
also means that every HiPERCAM run involves a significant amount of
extra staff time to mount/dismount the instrument at the start/end,
and risks damage to the instrument each time it is moved.

We have identified a solution to this problem -- the GTC has a third
Folded Cassegrain focus, labelled G, that has never been used. This
focus is currently just a hole in the steel structure of the telescope
through which the telescope beam can be steered by the tertiary
mirror. The focus currently has no image derotator (commonly referred
to as a {\em rotator}), cable wrap, autoguider, or services
(electricity, ethernet, coolant). The reason this focus has never been
commissioned by the GTC is that the surrounding space envelope is too
small to fit their common-user instrumentation. But this is not a
problem for HiPERCAM, which as a visitor instrument was designed to be
as compact as possible, and is far smaller than any of the other GTC
instruments.

We have recently completed a preliminary design of a compact rotator
for HiPERCAM that can fit in the available space envelope. One way in
which space has been saved is by not incorporating a traditional
autoguider mechanism with a probe arm mounted on an azimuthal track in
the rotator. Instead, autoguiding with the new rotator will be
provided in two ways. For high-speed observations (seconds and below),
guiding will be performed from the science images, as described in
Section~\ref{sec:pipeline}. For deep imaging, for which COMPO becomes
redundant, we shall use COMPO for autoguiding, with the pick-off arm
selecting guide stars outside the HiPERCAM field of view and the
injection arm redirecting the light to a separate autoguider camera
fixed to the interface collar on which COMPO is mounted (see
Section~\ref{sec:mech}).

We plan to begin manufacture of the new rotator in 2021, with
commissioning on the telescope during 2022.

\section{Conclusions}

We have presented the design of HiPERCAM and demonstrated that it is
performing to specification on the GTC. We have also described some of
the future upgrades planned for the instrument, including a novel
comparison-star pick-off system. HiPERCAM provides the GTC with a
unique capability amongst the world's 8--10\,m-class telescopes and is
a powerful new tool in the field of time-domain astrophysics.

\section*{Acknowledgments}

We thank the referee, Shrinivas Kulkarni, for his valuable comments.
HiPERCAM was funded by the European Research Council under the
European Union's Seventh Framework Programme (FP/2007-2013) under
ERC-2013-ADG Grant Agreement no. 340040 (HiPERCAM). Based on
observations made with the Gran Telescopio Canarias, installed
at the Spanish Observatorio del Roque de los Muchachos of the
Instituto de Astrof\'{i}sica de Canarias, on the island of La
Palma. We would like to thank the mechanical and electronics
technicians at Sheffield, UKATC and Durham for their major
contribution to the project. We would also like to thank the staff of
the ING and GTC for their assistance during commissioning, and
Guðmundur Stef\'{a}nsson and Suvrath Mahadevan for providing one of
their beam-shaping diffusers for testing in HiPERCAM. The many
discussions we had with the Spider team at Manchester (Rene Breton,
Colin Clark, Mark Kennedy, Daniel Mata S\'{a}nchez, Guillaume Voisin)
about HiPERCAM science data were invaluable in improving our
understanding of the instrument's performance. SGP acknowledges the
support of a Science and Technology Facilities Council (STFC) Ernest
Rutherford Fellowship. PR-G and TM-D acknowledge support from the
State Research Agency (AEI) of the Spanish Ministry of Science,
Innovation and Universities (MCIU), and the European Regional
Development Fund (FEDER) under grant AYA2017-83383-P. MAPT and TM-D
acknowledge support via Ram\'{o}n y Cajal Fellowships RYC-2015-17854
and RYC-2015-18148.

\section*{Data availability}

Only commissioning data are reported on in this paper. The data will
be shared on reasonable request to the corresponding author.

\bibliographystyle{mnras}
\bibliography{hicam_mn}

\bsp
\label{lastpage}
\end{document}